\documentclass[a4paper,12pt,twoside,openright]{UGthesis}
\usepackage{fancyhdr}
\usepackage[dvips]{graphicx}
\usepackage{cite}
\usepackage{amssymb}
\usepackage[mathscr]{eucal}
\usepackage[errorshow]{tracefnt}
\usepackage{amsmath}
\usepackage{amssymb}
\usepackage{amsthm}
\usepackage{eucal}
\usepackage{eufrak}

\newcommand{\vA}{\mathcal{A}}
\newcommand{\vB}{\mathcal{B}}
\newcommand{\vC}{\mathcal{C}}
\newcommand{\vD}{\mathcal{D}}
\newcommand{\vE}{\mathcal{E}}
\newcommand{\vF}{\ensuremath{\mathcal{F}}}
\newcommand{\vG}{\ensuremath{\mathcal{G}}}
\newcommand{\vH}{\ensuremath{\mathcal{H}}}

\newcommand{\vL}{\ensuremath{\mathcal{L}}}

\newcommand{\vN}{\ensuremath{\mathcal{N}}}
\newcommand{\vP}{\ensuremath{\mathcal{P}}}
\newcommand{\vQ}{\ensuremath{\mathcal{Q}}}

\newcommand{\vV}{\ensuremath{\mathcal{V}}}

\newcommand{\pa}{\partial}

\newcommand{\beq}{\begin{equation}}
\newcommand{\beqn}{\begin{equation*}}
\newcommand{\eeq}{\end{equation}}
\newcommand{\eeqn}{\end{equation*}}
\newcommand{\beqa}{\begin{eqnarray}}
\newcommand{\beqan}{\begin{eqnarray*}}
\newcommand{\eeqa}{\end{eqnarray}}
\newcommand{\eeqan}{\end{eqnarray*}}
\newcommand{\bdm}{\begin{displaymath}}
\newcommand{\edm}{\end{displaymath}}

\newcommand{\la}{\langle}
\newcommand{\ra}{\rangle}

\newcommand{\ba}{\begin{array}}
\newcommand{\ea}{\end{array}}

\newcommand\ffam{\sffamily}
\newcommand\fser{\bfseries}
\newcommand\fsh{\upshape}

\newcommand\nn{\nonumber}

\newcommand\benu{\begin{enumerate}}
\newcommand\eenu{\end{enumerate}}
\newcommand\bit{\begin{itemize}}
\newcommand\eit{\end{itemize}}

\def\End{\mathrm{End\,}}

\def\tr{\mathrm{tr\,}}
\def\ad{\mathrm{ad\,}}

\def\dim{\mathrm{dim\,}}

\def\max{\mathrm{max\,}}
\def\der'{\mathfrak{der}'\,}
\def\der{\mathfrak{der}\,}
\def\str'{\mathfrak{str}'\,}
\def\str{\mathfrak{str}\,}

\def\g{\gamma}

\def\H{\mathbb{H}}

\def\frake{\mathfrak{e}}

\def\g{\mathfrak{g}}
\def\frakp{\mathfrak{p}}
\def\u{\mathfrak{u}}
\def\so{\mathfrak{so}}
\def\su{\mathfrak{su}}
\def\sp{\mathfrak{sp}}
\def\sl{\mathfrak{sl}}
\def\gl{\mathfrak{gl}}

\newcommand{\h}{\mathfrak{h}}
\newcommand{\p}{\mathfrak{p}}
\newcommand{\frakk}{\mathfrak{k}}

\newcommand{\gjts}{generalized Jordan triple system }

\newcommand{\al}{\alpha}
\newcommand{\be}{\beta}
\newcommand{\ep}{\varepsilon}
\newcommand{\de}{\delta}
\newcommand{\ga}{\gamma}
\newcommand{\Ga}{\Gamma}

\newcommand{\da}{{\dot{\alpha}}}
\newcommand{\db}{{\dot{\beta}}}

\newcommand{\octic}{{\sffamily\bfseries Paper~II}}
\newcommand{\from}{{\sffamily\bfseries Paper~I}}
\newcommand{\gcrkma}{{\sffamily\bfseries Paper~III}}
\newcommand{\blgjts}{{\sffamily\bfseries Paper~IV}}
\newcommand{\gauging}{{\sffamily\bfseries Paper~V}}

\numberwithin{equation}{section}

\RequirePackage{hyperref}

\begin{document}

%
%



\pagenumbering{roman}
\setcounter{page}{1}

%
%

\thispagestyle{empty}




\begin{center}
{\upshape\sffamily\bfseries\huge 
\noindent
Exceptional Lie algebras \\[2mm] 
and M-theory}
 \\[4mm]
\end{center}

\begin{center}
        \rule{110mm}{2pt}
\end{center}

\vspace*{4mm}
\begin{center}
  {\fsh\ffam\fser\Large Jakob Palmkvist}\\
\end{center}

\begin{center}
  {\it Physique Th\'eorique et Math\'ematique\\
  Universit\'e Libre de Bruxelles \& International Solvay Institutes\\
  Boulevard du Triomphe, Campus Plaine, ULB-CP 231,\\BE-1050 Bruxelles, Belgium}\\
\end{center}

\begin{center}
  {\tt jakob.palmkvist@ulb.ac.be}\\
\end{center}

\begin{center}
  {\it Thesis for the degree of Doctor of Philosophy, defended on December 10, 2008,\\ at Fundamental Physics, Chalmers University of Technology, G\"oteborg, Sweden.\\
  The work was funded by the International Max Planck Research School for Geometric Analysis, Gravitation and String Theory, and conducted at the
Max Planck Institute for Gravitational Physics (Albert Einstein Institute) in Potsdam, Germany.}\\
\end{center}
\vspace*{2cm}

\centerline{\ffam\fser Abstract}
\medskip
\normalsize
\noindent
In this thesis we study algebraic structures in M-theory, in particular the exceptional Lie algebras arising in dimensional reduction of its low energy limit, eleven-dimensional supergravity. We focus on $\frake_8$ and its infinite-dimensional extensions $\frake_9$ and $\frake_{10}$. We review the dynamical equivalence, up to truncations on both sides, between eleven-dimensional supergravity and a geodesic sigma model based on the coset 
$E_{10}/K(E_{10})$, where $K(E_{10})$ is the maximal compact subgroup. The description of $\frake_{10}$ as a graded Lie algebra is crucial for this equivalence. We study 
generalized Jordan triple systems, which 
are closely related to graded Lie algebras, and which
may also play a role in the description of 
M2-branes using three-dimensional superconformal theories.

\newpage
%
%

\vspace*{4cm}

\centerline{\ffam\fser\Large Acknowledgments}
\medskip
\smallskip

\normalsize
\noindent
First of all, I would like to thank Professor Hermann Nicolai, not only in his role as my supervisor, but also as a director of the Albert Einstein Institute, where I have had the pleasure to work the last three years. I am also grateful to my official supervisor Professor Martin Cederwall and my examiner Professor Bengt E.~W.~Nilsson.
Furthermore, I would like to thank Axel Kleinschmidt
and
our collaborators in Groningen:
Eric A.~Bergshoeff, Olaf Hohm and Teake A.~Nutma, especially for their efforts to finish one of the papers when the deadline for my thesis was approaching.
I am grateful to Jonas Hartwig, Ling Bao and especially Daniel Persson
for carefully reading parts of the manuscript and giving me many valuable comments. I am also grateful to
Christoffer Petersson for generously lending me his sofa and his office during my visits in G\"oteborg. Finally, among the students and postdocs at the Albert Einstein Institute I would especially like to thank Sudarshan Ananth, 
Claudia Colonnello, Cecilia Flori, Thomas Klose, Michael Koehn, Carlo Meneghelli and Hidehiko Shimada for their support and friendship.

\newpage

\tableofcontents

\cleardoublepage

\renewcommand{\chaptermark}[1]{\markboth{Chapter \thechapter\ \ \ #1}{#1}}
\renewcommand{\sectionmark}[1]{\markright{\thesection\ \ #1}}
\lhead[\fancyplain{}{\sffamily\thepage}]%
  {\fancyplain{}{\sffamily\rightmark}}
\rhead[\fancyplain{}{\sffamily\leftmark}]%
  {\fancyplain{}{\sffamily\thepage}}
\cfoot{}
\setlength\headheight{14pt}


\setcounter{page}{5}
\pagenumbering{arabic}

%
%

\chapter{Introduction}

There are four fundamental forces in nature. Three of them, the electromagnetic, weak and strong interactions, can be described within the framework of quantum mechanics. The fourth force, gravity, is one that we all experience every day, but it is also the least understood of the four forces. Einstein's theory of general relativity works well in most situations, and it is already a great improvement of Newton's theory. However, at high energies and small distances, for example near the center of a black hole or shortly after the big bang, we need a quantum theory to describe gravity. In particular, this implies the existence of a spin two particle, called the graviton, mediating the 
force.

String theory was originally developed in the late 1960s as a theory of strong interaction, which keeps the quarks together within the hadrons. 
However, another description of strong interaction, called quantum chromodynamics (QCD) appeared in the early 1970s and turned out to be more successful. One of the drawbacks of string theory in this context is the existence of a spin two particle, which has no hadronic interpretation. However, this also has the advantage that string theory may, and indeed has to, be interpreted as a theory of quantum gravity. On the other hand, the spectrum of the bosonic string theory contains a tachyon, a state with negative mass squared.
One can get rid of this problem by imposing supersymmetry and considering superstrings instead of bosonic strings. 
Supersymmetry is a symmetry between bosons (particles that mediate forces) and fermions (particles that build up matter). 
Although not yet experimentally observed, supersymmetry is a very natural property to require for a theory of all known forces and matter, since it implies that the  
strengths of the electromagnetic, weak and strong interactions coincide at a certain energy scale.

A problem of bosonic string theory that cannot be solved by supersymmetry is the (natural) appearance of extra dimensions. Bosonic string theory does not work in the four-dimensional world that we live in, but requires 26 dimensions. Supersymmetry reduces this number, but only down to 10. One way to come around this obstacle is to think of some of the dimensions as closed circles instead of lines that are infinitely extended in both directions. If all except four of these circles are sufficiently small, they cannot be distinguished from points and the theory is effectively four-dimensional. This is an example of  
\textit{compactification} -- all but four of the dimensions are \textit{compact}. If we compactify $n$ dimensions, each spacetime point in the effective lower-dimensional theory can be interpreted as an $n$-dimensional manifold. In the example with a circle for each compact dimension, the resulting manifold is an $n$-torus, but there are other much more complicated possibilities. Compactification can be seen as a source of unification -- seemingly unrelated features of a theory can have a common origin in a higher-dimensional theory, compactified on an appropriate manifold.

In the first superstring revolution 1984--85 two new string theories in ten dimensions were found, called \textit{heterotic} string theories, with $SO(32)$ and $E_8 \times E_8$ as gauge groups, respectively. It was shown by Green and Schwarz that for these groups (but no others) all anomalies cancel \cite{Green:1984sg,Green:1984qs}. Moreover, upon compactification on a so called Calabi-Yau manifold the $E_8 \times E_8$ theory may lead to the gauge group $U(1) \times SU(2) \times SU(3)$ that describes the electromagnetic, weak and strong interactions. In addition to the heterotic theories, there were already two theories of closed strings, called type IIA and type IIB, and a fifth theory, called type I, with both open and closed strings.  

The fact that string theory on the one hand exhibited promising features as a theory of quantum gravity, and on the other hand required supersymmetry and extra dimensions, raised the interest in supergravity in various dimensions, and with various amount of supersymmetry. It was shown that eleven is the maximal number of dimensions for a supergravity theory with Minkowski signature and without particles of higher spin than two \cite{Nahm:1977tg}.   
Furthermore, in eleven dimensions there is only one supergravity theory \cite{Cremmer:1978km}, whereas there are more possibilities in lower dimensions. Dimensional reduction of eleven-dimensional supergravity on a circle gives type IIA supergravity, which is the low energy limit of type IIA string theory. More generally, reduction on an $n$-torus, 
gives maximal supergravity in $11-n$ dimensions.

In the second superstring theory revolution 1994--95, Hull, Townsend 
\cite{Hull}
and Witten \cite{Witten:1995ex} showed that the five string theories are connected by dualities. It was proposed that eleven-dimensional supergravity is the low energy limit of a more fundamental theory, called M-theory.
Unlike strings, the fundamental objects in M-theory are believed to be extended in not only one but two spatial directions. Such objects are called supermembranes or M2-branes.
Very little is known about M-theory but we can learn more about it by studying its low energy limit, eleven-dimensional supergravity, and its reductions.

Toroidal reduction of eleven-dimensional supergravity to $d=11-n$ dimensions gives rise to symmetries in the reduced theories, which are said to be \textit{hidden} since some of the fields must be dualized to make the symmetry manifest. After dualization the scalars in the $d$-dimensional theory parameterize the coset $G/K(G)$, where $G$ is the global symmetry group of the Lagrangian, and $K(G)$ is its maximal compact subgroup (which appears as a local symmetry).
For $4 \leq n \leq 8$, the symmetry groups $G$ are the exceptional groups $E_{n}$, with Lie algebras $\frake_n$ \cite{Cremmer:1979up,Cremmer:1997ct,Cremmer:1999du}.

In three dimensions all the bosonic degrees of freedom can be dualized to scalars and can thereby be described by a sigma model based on the coset 
$E_8/(\text{Spin}(16)/\mathbb{Z}_2)$ \cite{Julia4,Marcus:1983hb}. 
The fact that scalars are dual to scalars in two dimensions makes the
step from $d=3$ down to $d=2$ different from the preceding steps in the successive reduction.
The corresponding $E_9$ and $K(E_9)$ symmetries are not realized on the action but on the equations of motion, which can be written as an integrability condition of a linear system \cite{Nicolai:1987kz}.
This difference is on the mathematical side reflected by the fact that $\frake_9$ is infinite-dimensional, unlike $\frake_n$ for $4 \leq n \leq 8$. 
The appearance of infinite-dimensional symmetries in $d=2$ was first studied by Geroch for pure gravity reduced from four to two dimensions
\cite{Geroch:1970nt,Geroch:1972yt}.

One might suspect that $\mathfrak{e}{}_{10}$ should appear in the reduction to only one (time) dimension, or even $\mathfrak{e}{}_{11}$ in zero dimensions
\cite{Julia3,Julia2}. Partial results concerning $\frake_{10}$ were found in
\cite{Mizoguchi:1997si}.
Although $\frake_{9}$ and $\frake_{10}$ both are infinite-dimensional and both can be defined recursively, there is a crucial difference in complexity. For $\frake_9$, which is an \textit{affine} algebra, there is a pattern that repeats itself and makes it possible to write down all the commutation relations in a closed form. For $\frake_{10}$, a \textit{hyperbolic} algebra, the number of new elements grows exponentially for each step in the recursive definition, and soon one looses control over the algebra.

Beside the conjectural symmetry in the reduction to one dimension,
hyperbolic Kac-Moody algebras were also shown to appear near spacelike singularities in supergravity theories
\cite{Damour:2000hv,Damour:2001sa}. 
The chaotic behavior in this limit \cite{BKL} can be reformulated as
a billiard motion in the Weyl chamber of a hyperbolic Kac-Moody algebra, which for 
eleven-dimensional supergravity is $\frake_{10}$.

Inspired by the coset symmetries in dimensional reduction and the appearance of hyperbolic algebras in cosmological billiards,
Damour, Henneaux and Nicolai considered a one-dimensional geodesic sigma model based on the infinite-dimensional coset $E_{10}/K(E_{10})$ \cite{Damour:2002cu}. They found a correspondence, up to truncations on both sides, between the sigma model equations of motion and those of eleven-dimensional supergravity at a fixed, but arbitrarily chosen spatial point \cite{Damour:2002cu,Damour:2004zy}.
Corresponding results for the maximal supergravity theories in ten dimensions were obtained in \cite{Kleinschmidt:2004dy,Kleinschmidt:2004rg} using the same coset model, but different level decompositions.
The model has also been extended to the fermionic sector of eleven-dimensional supergravity, involving spinor and vector-spinor representations of $\mathfrak{k}(\frake_{10})$ \cite{deBuyl:2005zy,Damour:2005zs,Buyl:2005mt,Damour:2006xu}. These representations are finite-dimensional and thus \textit{unfaithful}, since the algebra itself is infinite-dimensional. There are problems with the model related to this fact, and the construction of a faithful fermionic representation would probably be an important progress. In an alternative 
approach it has been has been proposed that eleven-dimensional supergravity is a nonlinear realization of the Lorentzian algebra $\frake_{11}$ \cite{West:2001as}.
See also \cite{Englert:2003py,Englert:2004it} for a model combining the approaches in \cite{Damour:2002cu} and \cite{West:2001as}.

\newpage

\section{Outline}

This text consists of 
six chapters and is intended to be an introduction to the five 
research papers
\cite{Kleinschmidt:2006dy,Cederwall:2007qb,Palmkvist:2007as,
Nilsson:2008kq,Bergshoeff:2008xv}.

In chapter 2 we review how dimensional reduction gives rise to coset symmetries. We do this in detail for pure gravity in $D$ dimensions reduced to $d=D-n$ dimensions. We also discuss very briefly how the symmetry gets enhanced from 
$GL(n)$ to $SL(n+1)$
in $d=3$. Taking all the bosonic fields in supergravity into account, the global symmetry groups are extended to (the split real forms of) the exceptional groups $E_n$ for $4\leq n \leq 8$.
In order to describe the corresponding Lie algebras, in particular for $E_8$ and its infinite-dimensional extensions $E_9$ and $E_{10}$, we need the mathematical background presented in chapter 3 and 4.
The first of these chapters provides the standard classification of Kac-Moody algebras, including also the simple finite-dimensional Lie algebras (defined over the complex numbers). In the end of that chapter we extend the discussion to graded Lie algebras in general. The gradings of a Kac-Moody algebra, and the concomitant level decompositions of its adjoint representation are important, in particular in the infinite-dimensional cases where this is the only way to extract information that we can compare to physics. 

In chapter 4 we discuss generalized Jordan triple systems. These are algebraic structures that, on certain conditions on both sides, are in one-to-one correspondence with graded Lie algebras. 
We refine this general result to some special cases of graded Lie algebras and generalized Jordan triple systems 
that we are interested in. We call them \textit{nicely graded} Lie algebras and \textit{normed} triple systems. The nicely graded Lie algebras include the Kac-Moody algebras that appear in supergravity but also infinite-dimensional algebras that are not of Kac-Moody type. We explain how the corresponding normed triple systems are proposed to describe multiple M2-branes in three-dimensional superconformal theories.
In chapter 4 we also present the main result of \cite{Palmkvist:2007as} in a somewhat different formulation. Given two graded Kac-Moody algebras, such that one of their Dynkin diagrams is embedded in the other in a certain way, we show how the corresponding triple systems are related to each other.

In chapter 5 we study the exceptional algebras $\mathfrak{e}_n$, in particular
for $n=8,\,9,\,10$, and their maximal compact subalgebras $\mathfrak{k}(\frake_n)$. Many of the
results for 
$\frake_8,\,\frake_9,\,\frake_{10}$ hold in general for finite, affine and hyperbolic Kac-Moody algebras, respectively.
We apply the results in chapter 4 to examine the levels in the level decomposition of $\mathfrak{e}_n$ under the $\mathfrak{a}{}_{n-1}$ subalgebra. For $\frake_9$ we relate the $\mathfrak{a}{}_{8}$ levels 
to the affine levels that appear in the current algebra construction of $\frake_9$. We also study the spinor- and vector-spinor representations of $\frakk(\frake_n)$ that arise naturally in the fermionic extension of the original $E_{10}$ coset model. For $\frake_{10}$ we apply the result about generalized Jordan triple systems and show how $\frake_{10}$ can be constructed in this way from $\frake_8$.
Finally, in chapter 6 we review briefly the dynamical equivalence between the $E_{10}/K(E_{10})$ coset model and eleven-dimensional supergravity, up to truncations on both sides. On the $\frake_{10}$ side we only keep the first two positive 
$\mathfrak{a}{}_{9}$ levels.

Beside the introductory text, the thesis also includes the five 
papers
\cite{Kleinschmidt:2006dy,Cederwall:2007qb,Palmkvist:2007as,
Nilsson:2008kq,Bergshoeff:2008xv}, henceforth referred to as {\sffamily\bfseries Paper I--V}.
In {\from} \cite{Kleinschmidt:2006dy} we study the spinor and vector-spinor representations of $\frakk(\frake_{10})$ appearing in the fermionic extension of the original $E_{10}$ coset model. We show that the restriction to the $\frakk(\frake_9)$ subalgebra gives the correct
R-symmetry transformations of the fermions in two-dimensional $N=16$ supergravity \cite{Nicolai:2004nv}. 
In {\octic} \cite{Cederwall:2007qb} we give an explicit expression for the primitive $E_8$ invariant tensor with eight symmetric indices, motivated by possible applications to 
U-duality in the presence of higher-derivative terms.
{\gcrkma} \cite{Palmkvist:2007as} contains the result about generalized Jordan triple systems that we already mentioned above. We show how two such triple systems, derived from two graded Kac-Moody algebras $\g$ and $\mathfrak{h}$ (where $\mathfrak{h}$ should not be confused with the Cartan subalgebra of $\g$) are related to each other if $\g$ is a certain extension of $\mathfrak{h}$. Together with the Kantor-Koecher-Tits construction, which associates a Lie algebra to any Jordan algebra, this implies that $\frake_8,\,\frake_9$ and $\frake_{10}$ (and further extensions) can be constructed in a unified way from the exceptional Jordan algebra, consisting of hermitian $3 \times 3$ matrices over the octonions.
(However, we do not do this explicitly in the paper.)
In {\blgjts} \cite{Nilsson:2008kq} we study generalized Jordan triple systems in the context of superconformal M2-branes. We show that the recently
proposed theories with six or eight supersymmetries
can be entirely expressed in terms of the graded Lie algebra associated to a generalized Jordan triple system. 
Finally, in {\gauging} \cite{Bergshoeff:2008xv} we return to the bosonic $E_{10}$ coset model, this time applied to gauged maximal supergravity in three dimensions. We show that the embedding tensor that describes the gauge deformation arises naturally as an integration constant.

\chapter{Eleven-dimensional supergravity and its reductions}
\label{hiddenchapter}

We start with a brief account of the bosonic sector of eleven-dimensional supergravity \cite{Cremmer:1978km}. We will then review how coset symmetries arise in dimensional reduction of gravity \cite{Cremmer:1979up,Cremmer:1997ct,Cremmer:1999du}.
A good introduction into the subject, which we partly follow, is \cite{Pope}.

\section{Eleven-dimensional supergravity}

The bosonic sector of eleven-dimensional supergravity 
consists of an elfbein $E_M{}^A$
and a gauge field
$A_{	MNP}$, which is totally antisymmetric in the three indices. 

The curved indices $M,\,N,\,\ldots$ 
are lowered with the metric $g_{MN}$,
and the flat indices $A,\,B,\,\ldots$ with
\begin{align}
\eta_{AB}&=(-+\cdots+).
\end{align}
Both curved and flat indices take the eleven values $0,\,1,\,\ldots,\,10$. We will denote the inverse of the elfbein by $E_{A}{}^M$. Thus the position of curved and flat indices keeps the notation unambiguous.  

The bosonic theory
is described by the
Lagrangian \cite{Damour:2004zy}
\begin{align} \label{elevendimsugraaction}
\vL&=E(R - \tfrac1{48}F_{MNPQ}F^{MNPQ})
+12^{-4}
\ep^{MNPQRSTUVWX}F_{MNPQ}F_{RSTU}A_{VWX},
\end{align}
where we have introduced the determinant $E$ of the elfbein and 
the field strength
\begin{align}
F_{MNPQ}&=4 \pa_{[M} A_{NPQ]}
\end{align}
of the gauge field 
$A_{MNP}$.
The curvature scalar $R$ can be obtained from the elfbein
via the coefficients of anholonomy
\begin{align}
\Omega_{AB}{}^{C}=
2E_{[A}{}^M E_{B]}{}^N \pa_M E_N{}^C,
\end{align}
the spin connection
\begin{align}
\omega_{ABC}=\tfrac12(\Omega_{ABC}
+\Omega_{CAB}
-\Omega_{BCA}),
\end{align}
and the Riemann tensor (without torsion)
\begin{align}
R_{ABCD}&=2E_{[A|}{}^M\pa_M \omega_{|B]CD}
+2\omega_{[A|C}{}^E\omega_{|B]ED}
+2 \omega_{[AB]}{}^E\omega_{ECD},
\end{align}
which finally gives
\begin{align}
R=\eta^{AC}\eta^{BD}R_{ABCD}.
\end{align}
We note that 
the Riemann tensor
$R_{ABCD}$
is antisymmetric within the pairs of indices $[AB]$ and $[CD]$ but symmetric under exchange of the pairs. 
The spin connection is antisymmetric in the last pair of indices, 
$\omega_{ABC}=-\omega_{ACB}$,
and $2 \omega_{[AB]C} = \Omega_{ABC}$.

The bosonic equations of motion that follow from the Lagrangian (\ref{elevendimsugraaction}) read \cite{Damour:2004zy}
\begin{align}
D_{A}F^{ABCD}&=\tfrac1{8\cdot 144}\ep^{BCDEFGHIJKL}F_{EFGH}F_{IJKL},\nn\\
R_{AB}&=\tfrac1{12}F_{ACDE}F_B{}^{CDE}-\tfrac1{144}\eta_{AB}F_{CDEF}F^{CDEF}.
\end{align}
From the fact that partial derivatives commute we have the Bianchi identity 
\begin{align}
D_{[A}F_{BCDE]}&=0.
\end{align}
We will come back to these equations in chapter \ref{e10modelchapter}, when we study the $E_{10}$ coset model.

\section{Dimensional reduction of pure gravity}
If we set the gauge field $A_{MNP}$ in eleven-dimensional supergravity to zero,
then we are left with
pure gravity
in eleven dimensions,
\begin{align}
\vL=ER.
\end{align}
Pure gravity has the same form in any dimension, so we can as well be general and consider the $D$-dimensional theory. 
Thus we let the indices $A$ and $M$ take $D$ values.
We will perform a dimensional reduction on a (spatial) $n$-torus to $d=D-n$ spacetime dimensions.
For this we split the $D$-dimensional spacetime indices as
\begin{align}
M
&\to (\mu,
\,m)
&&\text{(curved indices)}\nn\\
A
&\to (\al,
a)
&&\text{(flat indices)}
\end{align}
where $\mu,\,\ldots$ and $\al,\,\ldots$ are the $d$-dimensional spacetime indices, while $m,\,\ldots$ and $a,\,\ldots$
take $n=D-d$ values.
We will raise and lower all small latin indices 
with the $SO(n)$ invariant metric $\delta$. Flat greek indices will be raised and lowered with $\eta$.

We will use hats for the $D$-dimensional quantities.
Quantities without hats are defined in the same way as above, but with $d$-dimensional indices.
We parameterize the vielbein as
\begin{align}
\hat{E}{}_\mu{}^\be &= e^{p\varphi} E_\mu{}^\be,   &  
\hat{E}{}_\mu{}^b & =e^{q\varphi} E_m{}^bA_\mu{}^m, \nn\\
\hat{E}{}_m{}^\be &=0, &   \hat{E}{}_m{}^b &= e^{q\varphi}E_m{}^b,
\end{align}
where $p$ and $q$ are constants that we will fix later, in order to have the reduced theory on a convenient form. 
The idea of Kaluza-Klein reduction is to interpret $A_{\mu}{}^m$ and $\varphi$
as $m$ vector fields and a scalar field (called the dilaton) in $d$ dimensions. 
Furthermore, unlike general compactificaation, we neglect all dependence on the compact dimensions, and set $\pa_m=0$.

We choose 
the dilaton
such that the internal vielbein 
$E_m{}^a$
has determinant one.
For the inverse of the vielbein we get 
\begin{align}
\hat{E}{}_\al{}^\nu &=e^{-p\varphi} E_\al{}^\nu,   &  
\hat{E}{}_\al{}^n&=-e^{-p\varphi} E_\al{}^\nu A_\nu{}^n, \nn\\
\hat{E}{}_a{}^\nu &=0, &   \hat{E}{}_a{}^n &= e^{-q\varphi}E_a{}^n.
\end{align}
We introduce 
`flat' derivatives $\hat{\pa}{}_A$ and ${\pa}{}_\al$, for which we have
\begin{align}
\hat{\partial}_\al = \hat{E}_\al{}^\mu \partial_\mu &= e^{-p\varphi}{E}_\al{}^\mu\partial_\mu =e^{-p\varphi}\partial_\al.
\end{align}
Now we get the following coefficients of anholonomy,
\begin{align}
e^{p\varphi}\hat{\Omega}_{\al\be\ga}&=
\Omega_{\al\be\ga}+p\eta_{\be\ga}\pa_\al\varphi-p\eta_{\al\ga}\pa_\be\varphi,\nn\\
e^{p\varphi}\hat{\Omega}_{\al\be c}&=e^{(q-p)\varphi}F_{\al\be c},\nn\\
e^{p\varphi}\hat{\Omega}_{\ga ab}&=
E_a{}^m \pa_\ga E_m{}_b
+q\pa_\ga\varphi\delta_{ab}
,\nn\\   
\hat{\Omega}_{\al b}{}^\ga&=\hat{\Omega}_{ab}{}^\ga
=\hat{\Omega}_{ab}{}^c=0,
\end{align}
where we have introduced the field strength
\begin{align}
F_{\al\be}{}^m&=2\pa_{[\al}(E_{\be]}{}^\mu A_\mu{}^m), &
F_{\al\be}{}^a&=F_{\al\be}{}^mE_m{}^a.
\end{align}
We proceed with 
the spin connection,
\begin{align}
e^{p\varphi}\hat{\omega}{}_{\al\be\ga}&=\omega_{\al\be\ga} 
+2p\eta_{\al[\be}\pa_{\ga]}\varphi
& \hat{\omega}{}_{abc}&=0
,\nn\\
e^{p\varphi}\hat{\omega}{}_{\al\be c}&=\tfrac12 e^{(q-p)\varphi}F_{\al\be}{}_c
,& e^{p\varphi}\hat{\omega}{}_{c\al \be}&=-\tfrac12 e^{(q-p)\varphi}F_{\al\be}{}_c,\nn\\
e^{p\varphi}\hat{\omega}{}_{ab \ga}&= P_\ga{}_{ab}+q\delta_{ab}\pa_\ga \varphi
, &
e^{p\varphi}\hat{\omega}{}_{\ga ab}&=Q_\ga{}_{ab},
\end{align}
where we have decomposed the Maurer-Cartan form $\hat{E}{}_a{}^m \pa_\ga \hat{E}{}_m{}_b$
into its symmetric and antisymmetric parts,
\begin{align} \label{uppdelning}
\hat{E}{}_{a}{}^m \pa_\ga \hat{E}{}_m{}_{b}&=\tilde{P}{}_\ga{}_{ab}+Q_\ga{}_{ab}, &
\tilde{P}{}_\ga{}_{ab}&=\hat{E}{}_{(a|}{}^m \pa_\ga \hat{E}{}_m{}_{|b)},&
Q{}_\ga{}_{ab}&=\hat{E}{}_{[a|}{}^m \pa_\ga \hat{E}{}_m{}_{|b]},
\end{align}
and furthermore 
taken out the trace,
\begin{align} 
\tilde{P}{}_\ga{}_{ab} &= P_\ga{}_{ab} + q \delta_{ab} \pa_\ga \varphi, &
P_\ga{}_{aa}&=0, & \tilde{P}{}_\ga{}_{aa}&={nq}\pa_\ga \varphi.
\end{align}
It is now straightforward to compute the 
Riemann tensor. The result is
\begin{align}
e^{2p\varphi}\hat{R}{}_{\al\be\ga\de}&=
{R}{}_{\al\be\ga\de}+4p\eta_{\be\ga}D_\al D_\de \varphi
-2p^2\eta_{\al\ga}\eta_{\be\de}D^\ep \varphi D_\ep \varphi
+4p^2\eta_{\be\de}D_\al \varphi D_\ga \varphi\nn\\
&\quad\,-\tfrac12 e^{2(q-p)\varphi}(F_{\al\ga a} F_{\be\de a}
+F_{\al\be e} F_{\ga\de e}),\nn\\
e^{2p\varphi}\hat{R}{}_{\al\be\ga d}&=
e^{(q-p)\varphi}(
pD_\ga \varphi F_{\al\be d}-pD_\al\varphi F_{\be\ga d}
+p\eta_{\ga \al}D^\ep\varphi F_{\be \ep d}
\nn\\&\quad\quad\quad\quad\,
-\tfrac12 D_\ga F_{\al \be d}-\tfrac12F_{\al\be e}\tilde{P}{}_{\ga de}-F_{\al\ga e}\tilde{P}{}_{\be de}),\nn\\
e^{2p\varphi}\hat{R}{}_{ab\ga\de}&=
-\tfrac12 e^{2(q-p)}F_{\ga\ep a}F_{\de}{}^{\ep}{}_{b}
-2\tilde{P}_\ga{}_{ae}\tilde{P}_\de{}_{be},\nn\\
e^{2p\varphi}\hat{R}{}_{a\be c\de}&=
2p\tilde{P}{}_{\de ac}D_\be\varphi-p\eta_{\be\de}
\tilde{P}{}_{\ep ac}D^\ep\varphi
-D_\be\tilde{P}{}_{\de ac}-\tilde{P}{}_{\be ae}\tilde{P}{}_{\de ce}
\nn\\
&\quad\,
+\tfrac14 e^{2(q-p)\varphi}F_{\be\ep c}F_{\de}{}^{\ep}{}_{a},\nn\\
e^{2p\varphi}\hat{R}{}_{abc \de}&=
e^{(q-p)\varphi}\tilde{P}{}_{\ep ac}F_{\de}{}^{\ep}{}_{b},\nn\\
e^{2p\varphi}\hat{R}{}_{abcd}&=
-2\tilde{P}{}_{\ep ac}\tilde{P}{}^{\ep}{}_{bd}.
\end{align}
with implicit (anti-)symmetrizations on the right hand side. The covariant derivative $D_\al$ is defined by
$D_\al = \pa_\al +\omega_\al + Q_\al$.
From the Riemann tensor we get 
\begin{align}
\hat{R} &= \eta^{AC}\eta^{BD}\hat{R}{}_{ABCD}=
e^{-2p\varphi}[R-s(\pa_\al\varphi)(\pa^\al\varphi)
-2((d-1)p+nq)D^2\varphi\nn\\&\qquad\qquad\qquad\qquad\qquad\quad\ \ 
-\tfrac14 g_{mn}e^{2(q-p)\varphi}\varphi F_{\al\be}{}^m F^{\al\be}{}^n
- {P}{}_\al{}_{ab} {P}{}^\al{}_{ab}],
\end{align}
where we have set 
\begin{align}
s=(d-1)(d-2)p^2+2n(d-2)pq+n(n+1)q^2
\end{align} 
for convenience.
The determinants of the vielbeine $\hat{E}_M{}^A$ and ${E}_m{}^a$ are related to each other as
$\hat{E}=e^{(dp+nq)\varphi}E$,
so we get 
\begin{align}
\vL=\hat{E}\hat{R}&=e^{
r\varphi}E
[
R-s(\pa_\al\varphi)(\pa^\al\varphi)
-2((d-1)p+nq)D^2\varphi\nn\\&\qquad\quad\ \,
-\tfrac14 
e^{2(q-p)\varphi} F_{\al\be}{}_a F^{\al\be}{}_a
- {P_\al{}_{ab} P^\al{}_{ab}}]
\end{align}
where 
$r=((d-2)p+nq)$. 
We set $r=0$ (Einstein frame), so that we can neglect the 
$D^2\varphi$ term as a total derivative. Furthermore, we choose the standard normalization $s=1/2$ of the kinetic term. Thus 
the constants are fixed to
\begin{align}
q&=-\frac{(d-2)}{n}p, &
p&=\pm\sqrt{\frac{n}{2(d-2)(d-2+n)}},
\end{align}
and we end up with the Lagrangian
\begin{align} \label{enkellagrangian}
\vL&=E[R-\tfrac12 (\pa_\al\varphi)(\pa^\al\varphi)
-\tfrac14 
e^{2(q-p)\varphi} F_{\al\be}{}_a F^{\al\be}{}_a - {P}{}_\al{}_{ab} {P}{}^\al{}_{ab}].
\end{align}
This Lagrangian
is invariant globally under $GL(n)$ and locally under $SO(n)$. 
To understand what this means, we rewrite the Lagrangian in terms of matrices.
We let $V$ be the internal vielbein, which is an $n \times n$ matrix with determinant one and components
$V_{m a}=E_m{}^a$. Furthermore, we let $M$ be the symmetric $n \times n$ matrix $VV^t$ with components
\begin{align}
M_{mn}=(VV^t)_{mn}=V_{ma}V_{na}=E_m{}^a E_n{}^a.
\end{align}
Finally, we interpret $P_\alpha{}_{ab}$ and ${F}{}_{\al\be}{}_a$ as the components of a traceless symmetric $n\times n$ matrix $P_\alpha$
and an $n \times 1$ column matrix ${F}{}_{\al\be}$.
After a little algebra we find that
\begin{align} \label{roligaformen}
\tr(P_\al P^\al) = -\tr(\pa_\al M \pa^\al M^{-1}),
\end{align}
and the last two terms in the Lagrangian (to be multiplied with the overall factor $E$) can be written
\begin{align} 
-\tr(P_\al P^\al)-\tfrac14 e^{2(q-p)\varphi} F_{\al\be}{}^t VV^t F^{\al\be}=
\tr(\pa_\al M \pa^\al M^{-1})-F_{\al\be}{}^t M F^{\al\be}.
\end{align}
First we show that this part of the Lagrangian has a global $SL(n)$ symmetry. Consider the transformations
\begin{align}
V &\to LV, & F_{\al\be}&\to(L^t)^{-1}F_{\al\be},
\end{align}
where $L$ is a constant $n\times n$ matrix with determinant one. This means that we replace $V$ by $LV$
and $F$ by $(L^t)^{-1}F$ everywhere. Then $M$ and $M^{-1}$ transform as
\begin{align}
M &\to LML^t,  &   M^{-1} &\to (L^{-1})^t M^{-1} L^{-1}.
\end{align}
Since $L$ is constant and the trace is invariant under cyclic permutations, we get
\begin{align} 
\tr(\pa_\al M \pa^\al M^{-1}) \to \tr(\pa_\al M \pa^\al M^{-1}),
\end{align}
and it is also easy to see that
\begin{align}
F_{\al\be}{}^t VV^t F^{\al\be} \to F_{\al\be}{}^t VV^t F^{\al\be}.
\end{align}
The first two terms in the Lagrangian (\ref{enkellagrangian}) and the overall factor $E$ do not depend on $V$ or $F$, so it follows that the whole expression is invariant.
Consider now the transformation
\begin{align}
V &\to VJ, 
\end{align}
where $J$ is an orthogonal $n \times n$ matrix, leaving $F$ invariant. Then we get
\begin{align}
M = VV^t \to VJJ^tV^t = VJJ^{-1}V^t = M,
\end{align}
so the Lagrangian is invariant even if $J$ is not constant.
The set of all $n \times n$ matrices with determinant one form the Lie group $SL(n)$ under matrix multiplication, and the subgroup $SO(n)$ consists of all orthogonal $n \times n$ matrices. What we have shown is that the Lagrangian (\ref{enkellagrangian}) has a global $SL(n)$ symmetry and a local $SO(n)$ symmetry. Alternatively, this can be shown by considering infinitesimal transformations. Then one acts with matrices that belong to the 
corresponding Lie algebras instead,
$\sl(n)$ and its subalgebra $\so(n)$.
They consist of all traceless and all antisymmetric $n \times n$ matrices, respectively. 
Thus $P_{\al}$ itself is an element of the Lie algebra $\sl(n)$, but not of $\so(n)$, whereas $M$ is an element of the Lie group $SL(n)$. 
The antisymmetric part $Q_{\al}$ of the Maurer-Cartan form,
which dropped out of the Lagrangian, is an element of the Lie algebra $\so(n)$ as well as of $\sl(n)$. 
Upon inclusion of the dilaton, or the trace part of the Maurer-Cartan form, $SL(n)$ and $\sl(n)$ extends to $GL(n)$ and $\gl(n)$.

\section{Hidden symmetries}

The $GL(n)$ and $SO(n)$ symmetries that we studied in the preceding section are examples of manifest symmetries -- they were already suggested by the use of curved and flat indices. However, when $d=3$, the symmetry gets enhanced from $GL(n)$ to $SL(n+1)$, although all the expressions have the same form, independently of $d$. The reason is that a $p$-form in $d$ dimensions has the same number of degrees of freedom as a $(d-2-p)$-form. Any $p$-form is dual to a $(d-2-p)$-form and they can be interchanged by \textit{dualization}. We will not explain this in detail, but 
as we have seen, we get $n$ vector fields $A_\mu{}^m$ in addition to the scalars when we reduce pure gravity from $D$ to $d=D-n$ dimensions. But in three dimensions, vectors are dual to scalars. This means that in addition to the components of the matrix $P_\al$ and the dilaton $\varphi$, we get $n$ extra scalars after dualization, and there are no other bosonic degrees of freedom. The number of scalars is thus the number of components of the matrix $P_\al$ plus $1+n$.
A symmetric $n \times n$ matrix has
\begin{align}
\dim{\,\sl(n)}-\dim{\,\so(n)}=n^2 - \tfrac12 n(n-1) = \tfrac12 n(n+1)
\end{align}
independent components, so
the total number of scalars after dualization is
\begin{align}
\tfrac12 n(n+1) + n + 1 = \tfrac12 (n+1)(n+2) = \dim{\,\sl(n+1)}-\dim{\,\so(n+1)}.
\end{align}
A detailed study shows that the symmetry is indeed $SL(n+1)$ globally and $SO(n+1)$ locally. Thus the whole Lagrangian (\ref{enkellagrangian})
can in this case be written as three-dimensional pure gravity coupled to a nonlinear sigma model of the form (\ref{roligaformen}).
For $D=11$, the number $m$ of scalars after reduction to $d$ dimensions is given by the following table.
\begin{align*}
\begin{array}{|r|r|lr|}
\hline
d & m & \multicolumn{2}{c|}
{\textit{after dualization}}\\
\hline
11 & 0&&0\\
10 & 1&&1\\
9 &
3&&3\\
8 & 
6&&6\\
7 & 
10&&10\\
6 & 
15&&15\\
5 & 
21&&21\\
4 & 
28&&28\\
3 & 
36&+8&=44\\
\hline
\end{array}
\end{align*}
So far, we have considered pure gravity. We will now extend the discussion to the full bosonic sector of maximal supergravity. Except for the vielbein $E_{M}{}^A$, it also contains an antisymmetric three-form $A_{MNP}$. 
It gives rise to two-forms, one-forms and scalars in lower dimensions. If we in each dimension $d$ dualize all $p$-forms such that
$d-2-p=0$, then we get the following total number of scalars from the original three-form.
\begin{align*}
\begin{array}{|r|r|lr|}
\hline
d & m & \multicolumn{2}{c|}{\textit{after dualization}}\\
\hline
11 & 0&&0\\
10 & 0&&0\\
9 & 0&&0\\
8 & 1&&1\\
7 & 4&&4\\
6 & 10&&10\\
5 & 20&+1&=21\\
4 & 35&+7&=42\\
3 & 56&+28&=84\\
\hline
\end{array}
\end{align*}
We denote the Lie algebras corresponding to the global and local symmetries by $\g$ and $\frakk(\g)$, respectively. In the next chapter we will see
that $\frakk(\g)$ is the \textit{maximal compact subalgebra} of $\g$. They extend $\sl(n)$ and $\so(n)$ to more intricate Lie algebras, given in the table below \cite{Cremmer:1979up,Cremmer:1997ct,Cremmer:1999du}.
\begin{align*}
\begin{array}{|r|c|c|r|}
\hline
d & \g & 
\frakk(\g)&\dim{\g}-\dim{\frakk(\g)}\\
\hline
10 & \u(1)&\{0\}&1-0=1\\ 
9 & \sl(2)\oplus \u(1)&\so(2)&4-1=3\\
8 & \sl(3)\oplus \sl(2)&\so(3)\oplus\so(2)&11-4=7\\
7 & \sl(5)&\so(5)&24-10=14\\
6 & \so(5,5)&\so(5)\oplus\so(5)&45-20=25\\
5 & \frake_{6(6)}&\sp(4)&78-36=42\\
4 & \frake_{7(7)}&\su(8)&133-63=70\\
3 & \frake_{8(8)}&\so(16)&248-120=128\\
\hline
\end{array}
\end{align*}
As for pure gravity, we will not show this is in detail, but 
try to convince the reader by counting the degrees of freedom. For 
any $3\leq d \leq 10$, the dimension of the coset $\frakp=\g\ominus\frakk(\g)$, (which is the rightmost number in the last table)
coincides with the sum of the number of scalars after dualization (which are the numbers in the two previous tables).

There is a Lie algebra $\frake_n$ (with split real form $\frake_{n(n)}$) for any $n \geq 6$, not only for $n=6,\,7,\,8$ as in the table above. 
In fact, we will see that the algebras $\g$ for $d=6$ and $d=7$ can be considered as $\frake_n$ for $n=5$ and $n=4$, respectively.
It is therefore natural to expect that also $\frake_9,\,\frake_{10}$ and $\frake_{11}$ show up in the reduction to two, one or even zero dimensions. However, we cannot proceed in the same way after $d=3$, since scalars become dual to scalars in two dimensions. On the mathematical side, this difficulty is reflected by the fact that the Lie algebras $\frake_n$ are infinite-dimensional for $n\geq9$. We will see how one can handle this in 
chapter \ref{enchapter}, 
but first we need some more general background about Lie algebras.

\chapter{Lie algebras} \label{liechapter}

The simple finite-dimensional Lie algebras were classified by Cartan and Killing a long time ago. In this classification, $\frake_6,\,\frake_7,\frake_8$ are included as \textit{exceptional} Lie algebras (together with $\mathfrak{f}{}_4$ and $\g_2$). As we will see in this chapter, there is a natural way to extend the classification, such that also some infinite-dimensional Lie algebras can be included. In particular, Lie algebras $\frake_n$ can be defined as such \textit{Kac-Moody algebras} for any $n \geq 4$. When we talk about exceptional Lie algebras in this thesis, we refer to this generalized meaning. All Lie algebras that we consider are defined over the complex numbers if nothing else is stated.
For introductions to Lie algebras and their representations, we recommend 
\cite{Fuchs,Fulton}.

\section{Kac-Moody algebras}

The Cartan-Killing classification of simple finite-dimensional Lie algebras
is based on the assignment of a (unique) Cartan matrix to any such Lie algebra, which describes it completely. By relaxing one of the conditions that a Cartan matrix must satisfy, one obtains a much larger class of Lie algebras, called Kac-Moody algebras
\cite{Kac67,Kac68A,Kac68B,Moody67,Moody68,Moody69}.
We will henceforth talk about Cartan matrices in this generalized meaning. The correspondence between Cartan matrices and Kac-Moody algebras 
is one-to-one up to isomorphisms between Kac-Moody algebras and permutations of the index set labeling rows and columns in the Cartan matrix \cite{Peterson-Kac83}.

In section \ref{cartanmatrissection},
we will review the classification of Cartan matrices, following \cite{Kac}. 
Since there is
a one-to-one correspondence between
Cartan matrices and Kac-Moody algebras
this will then correspond to a classification of Kac-Moody algebras.
In section \ref{chevser} we will explain how a Kac-Moody algebra
is constructed from its Cartan matrix $A$ if $\det{A}\neq0$.

\subsection{Cartan matrices}\label{cartanmatrissection}

Let $A$ be an indecomposable square matrix with integer entries $A_{ij}$. If $A_{ii}=2$ along the diagonal (no summation) and 
$A_{ij}\leq0$ for $i\neq j$, with
\begin{align}
A_{ij} = 0 \Leftrightarrow A_{ji} = 0, 
\end{align}
then $A$ is called a \textbf{Cartan matrix}.

For any column matrix $a$, we write $a>0$ if all entries are positive, and 
$a<0$ if all entries are negative.
We now define an $r \times r$ Cartan matrix $A$ to be
\begin{itemize}
\item
\textbf{finite} \textit{if $Ab > 0$,}
\item
\textbf{affine} \textit{if $Ab = 0$,}
\item
\textbf{indefinite} \textit{if $Ab < 0$}
\end{itemize}
for some $r\times 1$ matrix $b>0$.
One and only one of these three assertions is valid for any $A$, and  
in the affine case, $b$ is uniquely defined up to normalization \cite{Vinberg2,Kac}. 
Affine Cartan matrices can also be characterized in the following way.
\begin{itemize}
\item \textit{$A$ is affine if and only if $\det{A}=0$ and deletion of any row and the corresponding column gives a direct sum of finite Cartan matrices.} 
\end{itemize}

As we will describe in section \ref{chevser}, 
any Cartan matrix defines uniquely a Lie algebra, and all Lie algebras that can be obtained in this way are called Kac-Moody algebras.
Thus we can say that a Kac-Moody algebra is \textbf{finite}, \textbf{affine} or \textbf{indefinite} if the same holds for its Cartan matrix. Finite Kac-Moody algebras are then nothing but simple finite-dimensional Lie algebras, 
and their construction gives us back the
Cartan-Killing classification.
Also the affine Kac-Moody algebras are well understood, as certain extensions of finite algebras. 
On the other hand, the indefinite Kac-Moody algebras are neither fully classified
nor well understood.
We need to impose further conditions in order to study them along with 
the finite and affine algebras.
In what follows we will always require an indefinite Cartan matrix $A$ to be
\textbf{symmetrizable}, which means that 
there is a diagonal matrix $D$ with positive diagonal entries such that
$DA$ is symmetric. 
Then $D$ is unique up to an overall factor.
All finite and affine algebras are already symmetrizable \cite{Kac,Vinberg2}. 
It now follows that
\begin{itemize}
\item \textit{A is finite if and only if $A$ is symmetrizable and the symmetrized matrix has signature $(+\cdots++)$,}
\item \textit{A is affine if and only if $A$ is symmetrizable and the symmetrized matrix has signature $(+\cdots+\,0\,)$.}
\end{itemize}
\noindent
Analogously, we define $A$ to be \textbf{Lorentzian} if $A$ is symmetrizable and the symmetrized 
matrix has signature $(+\cdots+-)$. (With signature we mean the number of positive, negative or zero eigenvalues. Their order does not matter.) Clearly, the Lorentzian algebras form a subclass of the class of indefinite algebras, but we can restrict it even further.
Similarly to the characterization of the affine case above, we define 
\textbf{hyperbolic} Cartan matrices in the following way.
\begin{itemize}
\item \textit{$A$ is hyperbolic if and only if $\det A<0$ and deletion of any row and the corresponding column gives a direct sum of affine or finite matrices.}
\end{itemize}
It can be shown that any hyperbolic Cartan matrix is Lorentzian.
We say that a Kac-Moody algebra is \textbf{Lorentzian} or \textbf{hyperbolic} if the same holds for its Cartan matrix.

\subsection{Dynkin diagrams}

To any $r \times r$ Cartan matrix $A$, we can associate a 
graph which consists of $r$ nodes (labeled $1,\,2,\,\ldots,\,r$) and for each pair of nodes $(i,\,j)$ a number
$(\max\{|A_{ij}|,\,|A_{ji}|\})$ of lines between them. If the Cartan matrix $A$ is symmetric, then
this graph specifies it completely.
Such a graph, which contains all information about the Cartan matrix, is called the \textbf{Dynkin diagram} of the Cartan matrix.
However, if $A_{ij} \neq A_{ji}$ for some pair $(i,\,j)$, then 
the graph needs to be supplemented with additional information
in order to specify the Cartan matrix completely as
a Dynkin diagram. 
In the most interesting cases
this can be done by equipping the lines between the nodes $i$ and $j$ with an arrow, pointing towards $j$ if $|A_{ij}| > |A_{ji}|$.
In this thesis, we will mainly consider simply-laced algebras, which means that we only have the 
two possibilities below for the off-diagonal entries in the Cartan matrix (we recall that the diagonal entries are all equal to 2). 

\newcommand{\etioiiaapa}{
\begin{picture}(105,30)(-5,-10)
\thicklines
\multiput(10,0)(65,0){2}{\circle{5}}
\put(12.5,0){\line(1,0){60}}
\put(10,10){{\scriptsize$i$}}
\put(75,10){{\scriptsize$j$}}
\end{picture}}

\newcommand{\etioiibapa}{
\begin{picture}(105,30)(-5,-10)
\thicklines
\multiput(10,0)(65,0){2}{\circle{5}}
\put(10,10){{\scriptsize$i$}}
\put(75,10){{\scriptsize$j$}}
\end{picture}}

\newcommand{\iiatextapa}{
\begin{picture}(50,40)(-10,-10)
\put(-20,0){$A_{ij}=A_{ji}=0$}
\end{picture}}

\newcommand{\iibtextapa}{
\begin{picture}(50,40)(-10,-10)
\put(-20,0){$A_{ij}=A_{ji}=-1$}
\end{picture}}

\begin{align*}
\begin{array}{c|cc}
\!\!\!\!\!
\text{Dynkin diagram} 
\ &
\multicolumn{2}{l}{
\quad \,\,\,\,\text{Cartan matrix$\,\,$}}\\&&\\
\hline
\etioiibapa \  &\, &\iiatextapa\\
\etioiiaapa \  &\, &\iibtextapa\\
\end{array}
\end{align*}

\noindent
Since there is a one-to-one correspondence between Cartan matrices and Dynkin diagrams, we can talk about finite, affine and indefinite Dynkin diagrams.
The characterizations above of affine and hyperbolic matrices can now be formulated as
\begin{itemize}
\item \textit{$A$ is affine if $\det{A}=0$ and deletion of any node gives finite diagrams.} 
\item \textit{$A$ is hyperbolic if $\det A<0$ and deletion of any node gives affine or finite diagrams.}
\end{itemize}

Permutation of rows and (the corresponding) columns in $A$ corresponds to relabeling the nodes in the Dynkin diagram. The condition that a Cartan matrix should be indecomposable corresponds to the condition that a Dynkin diagram should be connected.

\subsection{The Chevalley-Serre relations} \label{chevser}

We will now describe how a Lie algebra can be constructed from a given Cartan matrix $A$ or, equivalently, from its Dynkin diagram. The Lie algebra $\g{}'$ obtained in this way is called  the \textbf{derived Kac-Moody algebra} of $A$. The {\it Kac-Moody algebra} $\g$ of $A$ is then defined as a certain extension of $\g{}'$ in the case when $A$ is affine. We will explain this in chapter
\ref{affinechapter}. If $A$ is finite or indefinite, then $\g$ coincides with $\g{}'$. 

In the construction of the Lie algebra $\g{}'$ from its Cartan matrix $A$, 
one starts with $3r$ generators $e_i,\,f_i,\,h_i$ satisfying the \textbf{Chevalley relations}
(no summation)
\begin{align} \label{chevalley-rel}
[e_i,\,f_j] &= \delta_{ij}h_j, & [h_i,\,h_j]&=0,\nn\\
[h_i,\,e_j] &= A_{ji}e_j,& [h_i,\,f_j] &= -A_{ji}f_j.
\end{align}
The elements $h_i$ span the abelian \textbf{Cartan subalgebra} $\h$.
The derived Kac-Moody algebra $\g{}'$ is then generated by $e_i$ and $f_i$
modulo the \textbf{Serre relations}
(no summation)
\begin{align} \label{serre-rel}
({\text{ad }e_i})^{1-A^{ji}}{e_j}&=0, & ({\text{ad }f_i})^{1-A^{ji}}{f_j}&=0.
\end{align}
It follows from the Chevalley relations (\ref{chevalley-rel}) that $\g$, except for the Cartan elements, is spanned by the set of multiple commutators
\begin{align} \label{rotvektormultkomm}
&[\cdots[[e_{i_1},\,e_{i_2}],\,e_{i_3}],\,\ldots ,\,e_{i_n}],&
&[\cdots[[f_{i_1},\,f_{i_2}],\,f_{i_3}],\,\ldots ,\,f_{i_n}],
\end{align}
for all $n \geq 1$,
which is restricted by the Serre relations (\ref{serre-rel}).
It also follows from (\ref{chevalley-rel}) that these multiple commutators 
are eigenvectors of $\text{ad }h$ for any $h \in \h$, and thus each of them defines an element $\mu$ in the dual space of $\h$, such that $\mu(h)$ is the corresponding eigenvalue. These elements $\mu$ are the \textbf{roots} of $\g$ and the eigenvectors are called \textbf{root vectors}. In particular, $e_i$ are root vectors of the \textbf{simple roots} $\alpha_i$, which form a basis of the dual space of $\h$. In this basis, an arbitrary root $\mu= \mu^i \alpha_i$ has integer components $\mu^i$, either all non-negative (if $\mu$ is a \textbf{positive root}) or all non-positive (if $\mu$ is a \textbf{negative root}). 

For {\it finite} Kac-Moody algebras, the space of root vectors corresponding to any root is one-dimensional. Furthermore, if $\mu$ is a root, then $-\mu$ is a root as well, but no other multiples of $\mu$. For any positive root $\mu$ of a finite Kac-Moody algebra $\g$, we let $e_\mu$
be a root vector corresponding to $\mu$ like the first one 
in $(\ref{rotvektormultkomm})$, with $\al_{i_1}+\al_{i_1}+\cdots+\al_{i_n}=\mu$.
In what follows we will write such a multiple commutator as
\begin{align} \label{e-rotvektormultkomm}
[e_{i_1},\,e_{i_2},\,\ldots ,\,e_{i_n}],
\end{align}
assuming the same ordering as in (\ref{rotvektormultkomm}).
Then $e_{\mu}$ is fixed up to a sign, and we let
$f_{\mu}$ be the root vector 
\begin{align} \label{f-rotvektormultkomm}
[-f_{i_1},\,-f_{i_2},\,\ldots ,\,-f_{i_n}],
\end{align}
corresponding to $-\mu$.
Thus a basis of $\g$ is formed by these root vectors $e_{\mu},\,f_\mu$ for all positive roots $\mu$, and by the Cartan elements $h_i$ for all $i=1,\,2,\,\ldots,\,r$. 

The reason for the minus signs in (\ref{f-rotvektormultkomm}) is that we 
can now also associate an element $h_\mu$ in the Cartan subalgebra
to each positive root in a way such that
the relations (\ref{chevalley-rel}) for $i=j$
extend from the simple roots to all positive roots,
\begin{align}
[e_\mu,\,f_\mu]&=h_\mu, & [h_\mu,\,e_\mu]&=2e_\mu, &
[h_\mu,\,f_\mu]&=-2f_\mu.
\end{align}
Now we can also 
define an involution $\omega$
on the finite Kac-Moody algebra $\g$ by
\begin{align}
\omega(h_\mu)&=-h_\mu, & \omega(e_\mu)&=-f_\mu, & \omega(f_\mu)&=-e_\mu,
\end{align}
for all positive rots $\mu=\mu^i \alpha_i$, where $h_\mu = \mu^i h_i$.
This involution is called the \textbf{Chevalley involution}.

For an arbitrary derived Kac-Moody algebra, 
there can be $m \geq 1$ linearly dependent root vectors to a given root, which is then said to have \textbf{multiplicity} $m$. Then the root vectors $e_{\mu}$ and $f_{\mu}$ are \textit{not} uniquely given by the root $\mu$ up to a sign, as for finite algebras. 
In order to distinguish between linearly independent root vectors corresponding to the same root,
we must also specify the order of the simple root vectors in the multiple commutators. 
The Chevalley involution is in this general case defined by
\begin{align}
\omega(h_i)&=-h_i, & \omega(e_i)&=-f_i, & \omega(f_i)&=-e_i
\end{align}
for the Chevalley generators, and then extended to the whole algebra by the homomorphism property.

\subsection{The Killing form}
In any finite-dimensional Lie algebra $\g$ we can define a bilinear form $\kappa$, called the \textbf{Killing form}, by
$\kappa(x,\,y)=\tr{(\ad{x} \circ \ad{y})}$.
The Killing form is symmetric and invariant under the adjoint action of the algebra,
\begin{align} 
\kappa([a,\,b],\,c)+\kappa(b,\,[a,\,c])=0.
\end{align}
Furthermore it is non-degenerate if and only if the Lie algebra is semisimple.
If $\g$ is simple and finite-dimensional then $\kappa$ can equivalently, up to an overall factor, be defined by
\begin{align}
\kappa(e_i,\,f_j)&=D_{ij}, &
\kappa(h_i,\,h_j)&=(DA)_{ij},
\end{align}
for all $i,\,j=1,\,2,\,\ldots,\,r$,
where $D$ is a diagonal matrix (unique up to an overall factor)
such that $DA$ is symmetric.
In all other cases the Killing form is defined 
to be zero. It can then be extended to the full algebra by the symmetry and invariance properties, together with the Chevalley relations. The Killing form will then be symmetric and invariant by construction, but also non-degenerate.
Moreover, these properties define the Killing form 
uniquely up to automorphisms and an overall normalization.

\subsection{The maximal compact subalgebra}
\label{maxkompsub}
The maximal compact subalgebra $\mathfrak{k}(\g)$ 
of a Kac-Moody algebra $\g$ is defined as the subalgebra of $\g$ which is pointwise fixed by the Chevalley involution.
Thus it consists of all elements $x+\omega(x)$, where $x \in \g$. 
A basis is given by $e_{\mu}-f_{\mu}$ for all root vectors $e_\mu,\,f_\mu$. 
Similarly, we define the coset $\mathfrak{p}(\g)$ 
to be the subspace consisting of all elements $x-\omega(x)$.
Thus it is spanned by all elements $e_{\mu}+f_{\mu}$, and the Cartan generators.
With respect to the Killing form,
the maximal compact subalgebra $\mathfrak{k}(\g)$ is negative-definite,
the coset $\mathfrak{p}(\g)$ is positive-definite away from the Cartan subalgebra,
and these two subspaces of $\g$
are orthogonal complements to each other. Moreover, we have
\begin{align} 
[\mathfrak{k},\,\mathfrak{k}] &\subset \mathfrak{k}, &
[\mathfrak{k},\,\mathfrak{p}] &\subset \mathfrak{p}, &
[\mathfrak{p},\,\mathfrak{p}] &\subset \mathfrak{k},
\end{align}
so the subspace $\p(\g)$ constitutes a representation of the subalgebra $\mathfrak{k}(\g)$, but $\p(\g)$ does not close under the Lie bracket.
The decomposition $\g=\mathfrak{k}+\mathfrak{p}$ is 
called the \textbf{Cartan decomposition}.

The maximal compact subalgebra $\mathfrak{k}(\g)$ of a Kac-Moody algebra $\g$ is itself a Kac-Moody algebra or a direct sum of Kac-Moody algebras as long as $\g$ is finite.
As we will see, this is in general not true if $\g$ is infinite-dimensional.

\subsection{Example: $\sl(n)$}
\label{chevgensln}
The Lie algebra $\mathfrak{a}_{n-1}$ ($n\geq 2$) has the following Dynkin diagram. 
\begin{center}
\scalebox{1}{
\begin{picture}(275,30)
\put(5,-10){$1$}
\put(45,-10){$2$}
\put(85,-10){$3$}
\put(230,-10){${n-1}$}
\thicklines
\multiput(10,10)(40,0){3}{\circle{10}}
\multiput(15,10)(40,0){2}{\line(1,0){30}}
\put(95,10){\line(1,0){15}}
\put(230,10){\line(1,0){10}}
\multiput(115,10)(10,0){12}{\line(1,0){5}}
\put(245,10){\circle{10}}
\end{picture}}\end{center}
\vspace*{0.4cm}
This Lie algebra can also be described as $\sl(n)$,  
consisting of all traceless $n\times n$ matrices
with the ordinary commutator as the Lie bracket.
It can be embedded into $\gl(n)$,  
the Lie algebra of all $n\times n$ matrices (not necessarily traceless).
As basis elements we can take all matrices $K^a{}_b$
for $a,\,b=1,\,2,\,\ldots,\,n$,
where the entry in row $a$ and column $b$ is one, and all other entries are zero. The commutation relations are then
\begin{align}
[K^a{}_b,\,K^c{}_d]=\de^c{}_bK^a{}_d-\de^a{}_dK^c{}_b.
\end{align}
The subalgebra $\sl(n)$ is obtained by factoring out the one-dimensional ideal spanned by the identity matrix.
Then we can write the Chevalley generators as
\begin{align} \label{chevgensln-ekv}
h_a &= {K^{a+1}}_{a+1}-{K^a}_a, & e_a &= {K^{a+1}}_a, & f_a &= {K^a}_{a+1}.
\end{align}
and we see that the Chevalley involution is given by minus the transpose. Thus the maximal compact subalgebra of $\sl(n)$ is $\so(n)$,
consisting of all antisymmetric matrices.
By embedding $\sl(n)$ into $\gl(n)$,
the Killing form can be written
\begin{align}
\kappa(K^a{}_b,\,K^c{}_d)=\de^c{}_b\de^a{}_d+m\de^a{}_b\de^c{}_d
\end{align}
for an arbitrary number $m$ (the terms involving $m$ cancel out for the $\sl(n)$ subalgebra). For two arbitrary traceless matrices $x$ and $y$, this gives $\kappa(x,\,y)=\tr{(xy)}$.

\section{Graded Lie algebras}

Kac-Moody algebras are special cases of \textit{graded} Lie algebras. In fact, it was the interest in graded Lie algebras that led Kac (independently of Moody) to the study of Kac-Moody algebras \cite{Kac}.

With a \textbf{graded} (or $\mathbb{Z}$-graded) Lie algebra we mean a Lie algebra that can be written as a direct sum of subspaces 
$\g_k \subset \g$ for all integers $k$, such that 
\begin{align} \label{gradering}
[\g_m,\,\g_n] \subseteq \g_{m+n}
\end{align}
for all integers $m,\,n$.
If there is a positive integer $m$ such that $\g_{\pm m} \neq 0$ but $\g_{\pm k} = 0$ for all $k > m$, then the Lie algebra $\g$ is $(2m+1)$-{graded}.
We will occasionally use the notation $\g_{\pm}=\g_{\pm 1}+\g_{\pm 2}+\cdots$.
A \textbf{graded involution} $\tau$ on the Lie algebra $\g$ is an automorphism such that
$\tau^2(x)=x$ 
for any $x \in \g$ and $\tau(\g_{k}) = \g_{-k}$ for any integer $k$.
A \textbf{characteristic element} \cite{Asano,Faraut} in a graded Lie algebra $\g$ is an element $d$ such that 
\begin{align}
[d,\,x] = kx
\end{align}
if $x \in \g_k$, for all integers $k$.
Any semisimple finite-dimensional graded Lie algebra has a characteristic element \cite{Kantor3.5}.

\subsection{Graded Kac-Moody algebras}

Consider a simple Kac-Moody algebra $\g$ and choose a simple root $\alpha_i$.
For any negative (positive) integer $k$, let $\g_k$ be the subspace of $\g$ spanned by all 
root vectors
$e_{\mu}$ ($f_{\mu}$) such that  
the component $\mu^i$ of the root $\mu$, corresponding to $\alpha_i$ in the basis of simple roots, is equal to $|k|$. Let $\g_0$ be spanned by 
all Cartan generators $h_j$ and all 
root vectors
$e_{\mu}$ and $f_{\mu}$ such that  
$\mu^i=0$.
In this way any simple root $\alpha_i$ gives a grading of $\g$.

Generally, any
set of simple roots ${\alpha}_{i_1},\,{\alpha}_{i_2},\,\ldots,\,{\alpha}_{i_n}$ gives a 
grading of $\g$ where $\g_k$ is spanned by all root vectors $e_{\mu}$ or $f_{\mu}$ such that
${\mu}^{i_1}+{\mu}^{i_2}+\cdots+{\mu}^{i_n}=\pm k$ and, if $k=0$, the Cartan generators.
Any grading of a simple finite-dimensional Lie algebra $\g$, such that
$[\g_i,\,\g_j]=\g_{i+j}$ for all integers $i,\,j$,
is given by a set of simple roots in this way
\cite{Kantor3.5}.
With a graded Kac-Moody algebra we will always mean a simple Kac-Moody algebra togehter with such a grading for some set $S$ of simple roots. It follows that the Chevalley involution is a graded involution in a graded Kac-Moody algebra.
Any graded involution $\tau$ together with the Killing form $\kappa$ on $\g$ induces a non-degenerate bilinear form on the subspace $\g_{-1}$, given by
\begin{align}
(x,\,y)=\kappa(x,\tau(y))
\end{align}
for all $x,\,y\in\g_{-1}$.
We call this the bilinear form \textbf{associated} to $\tau$.
A finite or indefinite graded Kac-Moody algebra has a unique characteristic element $d$ in the Cartan subalgebra. Its components in the basis of Cartan generators are given by the solution to the equation $Ad=b$
where $A$ is the Cartan matrix of $\g$ and
$b_i=1$ if $\alpha_i$ belongs to the set $S$ 
that defines the grading, and $b_i=0$ otherwise. 
Since $\g$ is simple, $\det{A}\neq0$ and the equation has a unique solution.
The subalgebra $\g_0$ is a direct sum of one-dimensional Lie algebras spanned by 
the Cartan elements corresponding to the set $S$ and
a direct sum $\g_0{}'$ of derived Kac-Moody algebras. 
Their Dynkin diagrams are obtained from that of
$\g$ by deleting the nodes that correspond to the set 
$S$ of simple roots.

\subsection{Level decomposition} \label{leveldecomp}

The grading of a Kac-Moody algebra comes with a \textbf{level decomposition} 
of its adjoint representation under the $\g_0{}'$ subalgebra. 
Since $[\g_0,\,\g_m] \subset \g_m$ as a special case of (\ref{gradering}), the subspace $\g_m \subset \g$ constitutes a representation ${\bf r}_m$ of $\g_0{}'$, which is the representation at level $m$ in the level decomposition. 
It follows that ${\bf r}_m$ is a subrepresentation of the $m$-fold tensor product $({\bf r}_1)^m$, irreducible or not. At level 2, only the antisymmetric part of ${\bf r}_1\otimes {\bf r}_1$ occurs, because of antisymmetry of the Lie bracket. At level 3, the totally antisymmetric part of 
${\bf r}_1\otimes {\bf r}_1\otimes {\bf r}_1$ is ruled out because of the Jacobi identity. In addition, the Serre relations restrict the representation at any nonzero level. 

\subsection{The universal graded Lie algebra}
\label{ugla-section}

In this section we will show how any vector space $V$ naturally gives rise to a graded Lie algebra
\begin{align}
\tilde{U}(V)=\tilde{U}_- + \tilde{U}_0
 + \tilde{U}_+.
\end{align}
As we will see, any graded Lie algebra $\g$ such that 
$[\g_{-m},\,\g_{-n}]=\g_{-m-n}$
for all positive integers $m,\,n$ can be embedded in 
$\tilde{U}(\g_{-1})$ \cite{Kantor-graded}. This will be important when we consider generalized Jordan triple systems in the next chapter. Moreover, any graded Lie algebra can be embedded into $\tilde{U}(\g_{-})$, which gives a nonlinear realization of $\g$ 
\cite{Kantor3.5}. 
A well known example of this is the conformal realization of $\so(2,\,d)$, as we explain in {\gcrkma}.

With an operator of order $p\geq 1$ on a vector space $V$ we mean a 
$p$-linear map $V^p \to V$. 
Let $A$ and $B$ be operators on a vector space $V$ of order $p$ and $q$, respectively. Then we define 
the composition $A \circ B$ to be an operator on $V$ of order $p+q-1$ by
\begin{align} \label{kantorkomposition}
&(A \circ B)(v_1,\,\ldots,\,v_{p+q-1})\nn\\
&\quad=\sum_{m=1}^p \sum A(v_{n_1},\,\ldots,\,v_{n_{m-1}},\,B(v_{n_m},\,
\ldots,\,v_{n_{m+q-1}}),\,v_{q+m},\,\ldots,\,v_{p+q-1})
\end{align}
where the second sum goes over all distinct values 
of the $q$ indices
$n_m,\,\ldots,\,n_{m+q-1}$ chosen among the 
$m+q-1$ values $1,\,\ldots,\,m+q-1$, such that
$n_m < \cdots < n_{m+q-1}$
and the remaining indices are ordered such that $n_1 < \cdots < n_{m-1}$.
For any integer $k \geq 0$, let $\tilde{U}{}_k$ be the vector space spanned by all operators
on $V$ of order $k+1$,
and let
$\tilde{U}{}_0 + \tilde{U}{}_+$ be the direct sum of all these vector spaces \cite{Kantor-graded}.
Then $\tilde{U}{}_0 + \tilde{U}{}_+$ is a graded Lie algebra under the commutator 
\begin{align} \label{kantorkommutator}
[A,\,B]= A\circ B-B\circ A.
\end{align}
A basis element $A$ in $\tilde{U}{}_{p-1}$ for $p\geq1$ is thus an operator on $V$ of order $p$, but it can also be viewed as a linear map
$V \to \tilde{U}{}_{p-2}$
by
\begin{align}
A(v_1)(v_2,\,\ldots,\,v_p) \equiv A(v_1,\,v_2,\,\ldots,\,v_p).
\end{align}
The vector spaces $\tilde{U}{}_k$ for $k\geq-1$ can be defined recursively in this way, starting from $\tilde{U}{}_{-1}=V$. We will in the continuation use this definition of $\tilde{U}{}_k$ instead of the one above by Kantor \cite{Kantor-graded}. One reason for this is that (\ref{kantorkomposition}) and (\ref{kantorkommutator}) then can be replaced by the simple formula
\begin{align}
[A,\,B]&=({\text{ad }A}) \circ B - ({\text{ad }B}) \circ A,
\end{align}
where, if $B$ is of order zero, $[A,\,B(v)]$ should be read as $A(B(v))$ for any $v$ in $V$.
It is straightforward to show by induction that the two definitions of the Lie bracket are equivalent. 

Having defined $\tilde{U}{}_+$ and $\tilde{U}{}_0$, we complete the vector space $\tilde{U}(V)$ by defining 
$\tilde{U}{}_-$ as the free Lie algebra generated by 
$V=\tilde{U}{}_{-1}$, 
where $\tilde{U}{}_{-k}$, for any $k\geq1$, is spanned by all multiple commutators
$[v_1,\,v_2,\,\ldots,\,v_k]$.
We extend the Lie algebra structure on the subspaces $\tilde{U}{}_-$
and $\tilde{U}{}_0 + \tilde{U}{}_+$ to the whole of $\tilde{U}(V)$ by
\begin{align}
[A,\,v]=A(v)
\end{align}
for any $A$ in $\tilde{U}{}_0 + \tilde{U}{}_+$ and any $v$ in $V$.
Since $\tilde{U}{}_-(V)$ is generated by elements in $V$,
this defines all
commutation relations between $\tilde{U}{}_-$
and $\tilde{U}{}_0 + \tilde{U}{}_+$ by the Jacobi identity.
The resulting algebra $\tilde{U}(V)$ is the \textbf{universal graded Lie algebra} 
of $V$ \cite{Kantor-graded}.

Any graded Lie algebra $\g$ such that $[\g_{-m},\,\g_{-n}]=\g_{-m-n}$
for all positive integers $m,\,n$ is isomorphic to the direct sum
\begin{align}
(\tilde{U}{}_-
/D) \oplus U_0
\oplus  U_+
\subset \tilde{U}(\g_{-1}),
\end{align}
where 
$U_0 \oplus  U_+$ is a certain subalgebra 
of $\tilde{U}{}_0 \oplus \tilde{U}{}_+$,
and $D$ is a graded ideal of $\tilde{U}{}_-$ 
\cite{Kantor-graded}. 
(This means that $D$ is the direct sum of its subspaces $D \cap U_k$ for all $k$.)

In fact, it is possible to embed any graded Lie algebra $\g$ in a universal graded Lie algebra $\tilde{U}(V)$, such that $\g$ is entirely contained in 
$\tilde{U}{}_{-1} +\tilde{U}{}_0 + \tilde{U}{}_+$. But then we must let the vector space $\g$ be the whole of $\g_-$, and not only $\g_{-1}$. Furthermore, the basis elements in $\g$ are mapped onto \textit{symmetric} operators in $\tilde{U}{}_0 + \tilde{U}{}_+$.
An operator $A$ on $V$ of order $p$ is \textbf{symmetric} if 
\begin{align}
A(\ldots,\,v_i,\,\ldots,\,v_j,\,\ldots)
= A(\ldots,\,v_j,\,\ldots,\,v_i,\,\ldots)
\end{align}
for all $1 \leq i,\,j \leq p$.
Then there is a corresponding map
$a:V \to V$ defined by
\begin{align}
a(v)=A(v,\,\ldots,\,v).
\end{align}
Conversely, $A$ is uniquely given by $a$, so we can identify $A$ with $a$ as an element in $\tilde{U}{}_{-1} +\tilde{U}{}_0 + \tilde{U}{}_+$. The composition of $a$ with another such element $b$ is a symmetric operator as well, given by 
\begin{align}
(a \circ b)(v)=p A(b(v),\,v,\,\ldots,\,v).
\end{align}
Let $M(V)$ be the subalgebra of $\tilde{U}{}_{-1} +\tilde{U}{}_0 + \tilde{U}{}_+$ spanned by all symmetric operators.
Then $M(V)$ is a graded Lie algebra with $M_{-k}=0$ for $k\geq -2$.

It has been shown \cite{Kantor3.5} (see also \cite{Kantor5}) that there is an injective homomorphism $\chi: \g \to M(\g_{-})$ given by
\begin{align} 
\label{kantor-formel}
\chi(u) : x \mapsto \left( \frac{{\text{ad }x}}{1-e^{-{\text{ad }x}}}\mathbb{P} e^{-{\text{ad }x}}\right)(u),
\end{align}
where $\mathbb{P}$ is the projection onto $U_-$ along $U_0 + U_+$, and the ratio should be considered as the power series
\begin{align} \label{potens-serie}
\frac{{\text{ad }x}}{1-e^{-\text{ad }x}}=
1+\frac{{\text{ad }x}}{2}+\frac{{(\text{ad }x)}^2}{12}-\frac{{(\text{ad }x)}^4}{720}+\cdots.
\end{align}
Since $\g$ is graded, $\chi$ induces a grading on $\chi(\g)$.
However, this is not the same grading as the one that $\chi(\g)$ is equipped with as a subalgebra of
$\tilde{U}{}_{-1} +\tilde{U}{}_0 + \tilde{U}{}_+$.

The grading that $\chi$ induces on $\chi(\g)$ can be defined on $M({V})$ for an arbitrary vector space $V$,
that is a direct sum of (infinitely many) 
subspaces $V_1,\,V_2,\,\ldots$.
Let $A$ be an element in $M(V)$ (that not necessarily belongs to one of the subspaces $M_k$). Then we can write $A(v)$ as a sum of $A_n(v)$ for all 
$n=1,\,2,\,\ldots$, where $A_n$ is a map $V \to V_n$. Suppose that
$A_n$ is a $(p_1 + p_2 + \cdots)$-linear map
\begin{align}
A_n : (V_1)^{p_1} \times (V_2)^{p_2} \times \cdots \to V_n,
\end{align}
symmetric under permutation of elements that belong to the same vector space $V_m$.
(In general, $A_n$ will be a sum of such maps.)
Then we say that $A$ has {grade} $p$ if
$p_1+ 2p_2 + 3p_3 + \cdots = n+p$ for all $n$.
Note that the grade can also be negative.

As before, we can identify $A$ as a symmetric operator of grade $p$ with a corresponding map $a : V \to V$. Then the 
composition $a \circ b$ of $a$ and another symmetric operator 
$b$, of grade $q$, is now the symmetric operator of grade $p+q$ given by 
\begin{align}
(a \circ b)_n (v) &= p_1A_n(b(v)_1,\,v_1,\,\ldots,\,v_1;\,v_2,\,v_2,\,\ldots,\,v_2;\ldots)\nn\\&\quad +p_2A_n(v_1,\,v_1,\,\ldots,\,v_1;\,b(v)_2,\,v_2,\,\ldots,\,v_2;\ldots)
+\cdots
\end{align}
for all $n=1,\,2,\,\ldots$, and a Lie bracket as usual by 
$[a,\,b]=a \circ b - b \circ a$. 
It follows that $M(V)$ is a graded Lie algebra also with this grading, which 
is preserved by the inverse of the homomorphism $\chi$ for $V=\g_{-1}$.

\chapter{Generalized Jordan triple systems} \label{gjtschapter}

In the end of the preceding chapter, we saw that any vector space $V$ gives rise to a graded Lie algebra $\tilde{U}(V)$.
Furthermore, we said that
any graded Lie algebra $\g$ 
generated by its $\g_{\pm 1}$ subspaces can be embedded in $\tilde{U}(V)$ 
for some vector space $V$. In this chapter we will see how a given graded Lie algebra can be extracted from $\tilde{U}(V)$. This is done by
identifying $V$ with $\g_{-1}$ and adding extra structure to this vector space, corresponding to the properties of $\g$.
The result is a \textit{generalized Jordan triple system}. 

In this chapter we will study various kinds of generalized Jordan triple systems and their associated graded Lie algebras. 
One reason for this is that we might learn more about the exceptional Lie algebras appearing in supergravity by studying their corresponding generalized Jordan triple systems. We can also go in the opposite direction and learn more about a generalized Jordan triple system by studying its associated graded Lie algebra.
In the end of this chapter we will explain how a certain kind of generalized Jordan triple systems, called three-algebras, are used in three-dimensional superconformal theories to describe multiple M2-branes.
In {\blgjts} we show how the theory can equivalently be formulated in terms of the associated graded Lie algebra. Such a formulation might lead to a generalization of the theory, since $\g_{-1}$ (the three-algebra) is only one of many subspaces of a graded Lie algebra $\g$.

\section{Preliminaries}
A {\bf triple system} is a vector space $V$ together with a trilinear map
\begin{align}
V \times V \times V &\to V, & (x,\,y,\,z) &\mapsto (xyz),
\end{align}
called {\bf triple product}.
Let $\g$ be a graded Lie algebra
with a graded involution $\tau$.
Then $\g_{-1}$ 
is a triple system with the triple product
\begin{align}
(xyz)= [[x,\,\tau(y)],\,z]. \label{standardtrippelprodukten}
\end{align} 
As a consequence of the Jacobi identity and the fact that $\tau$ is an involution, this triple product satisfies the identity
\begin{align}
(uv(xyz))-(xy(uvz))=((uvx)yz)-(x(vuy)z). \label{pre-GJTS-identity}
\end{align}
Any triple system that satisfies (\ref{pre-GJTS-identity})
is called a \textbf{generalized Jordan triple system}. 

Let $V$ be an arbitrary \gjts$\,$(not necessarily derived from a Lie algebra as above) 
of dimension $m$, and let $T^{\vA}$ be a basis of $V$, for $\vA=1,\,2,\,\ldots,\,m$. In analogy with Lie algebras we introduce \textbf{structure constants} $f^{\vA\vB\vC}{}_\vD$ for $V$, which specify the triple product by
\begin{align}
(T^\vA T^\vB T^\vC) = f^{\vA\vB\vC}{}_\vD T^\vD.
\end{align}
The identity (\ref{pre-GJTS-identity}) can then be written
\begin{align}
f^\vA{}^\vB{}^\vC{}_\vD f^\vE{}^\vF{}^\vD{}_\vG-
f^\vE{}^\vF{}^\vC{}_\vD f^\vA{}^\vB{}^\vD{}_\vG=
f^\vE{}^\vF{}^\vA{}_\vD f^\vD{}^\vB{}^\vC{}_\vG-
f^\vF{}^\vE{}^\vB{}_\vD f^\vA{}^\vD{}^\vC{}_\vG.
\label{generalized Jordan triple systemid}
\end{align}
\section{Normed triple systems}
Suppose now that $\g$ has a non-degenerate bilinear form $\kappa$, which is symmetric and invariant,
\begin{align}
\kappa(x,\,y)&=\kappa(y,\,x),&
\kappa([x,\,y],\,z)&=\kappa(x,\,[y,\,z]), \label{invarians-ejkm}
\end{align}
and such that
$\kappa(\g_m,\,\g_n)=0$
whenever $m+n\neq0$. The obvious example of such a bilinear form is the Killing form in a graded Kac-Moody algebra, but we want to
be more general here and
allow for Lie algebras that are not of Kac-Moody type.
Together with the involution $\tau$, the bilinear form $\kappa$
on $\g$ induces a bilinear form on $\g_{-1}$ by
\begin{align} \label{bilformen}
h(x,\,y)=\kappa(x,\,\tau(y)).
\end{align}
We call $h$ the bilinear form \textbf{associated} to $\tau$.
Suppose that $h$ is symmetric 
(which means that $\kappa$ is preserved by $\tau$).
Then we call $\g$ a \textbf{nicely graded} Lie algebra. 
As a consequence of the invariance (\ref{invarians-ejkm}), we have
\begin{align}
h(w,\,(xyz))=h(y,\,(zwx))=h(x,\,(wzy))=h(z,\,(yxw)). \label{normedgjts}
\end{align}
If this identity holds for some symmetric bilinear form $h$ defined on a 
generalized Jordan triple system $V$, then we say that $h$ is a \textbf{metric} on $V$, and that $V$ is a \textbf{normed}
triple system. Thus any nicely graded Lie algebra gives rise to a normed triple system.
We introduce the components $h^{\vA\vB}$ of the metric by
\begin{align}
h^{\vA\vB}=h(T^\vA,\,T^\vB).
\end{align}
We use $h^{\vA\vB}$ and the inverse $h_{\vA\vB}$ to raise and lower indices, for example,
\begin{align}
T_\vA &= h_{\vA\vB}T^\vB, & f^{\vA\vB\vC\vD}&= f^{\vA\vB\vC}{}_{\vE}h^{\vD\vE}.
\end{align}
The identity (\ref{normedgjts}) can now be written
\begin{align} \label{indexflytt}
f^{\vA\vB\vC\vD}=f^{\vC\vD\vA\vB}=f^{\vB\vA\vD\vC}=f^{\vD\vC\vB\vA}.
\end{align}

\section{The associated Lie algebra}

\label{lagjtssection}

In section \ref{ugla-section}, we defined the universal graded Lie algebra 
$\tilde{U}(V)$ of an arbitrary vector space $V$.
We will now assume that $V$ is a generalized Jordan triple system, and use this to define a subalgebra of $\tilde{U}(V)$.

For any pair of basis elements $(T^\vA,\,T^\vB)$ of $V$, we define the linear map
\begin{align}
S^{\vA\vB}:V \to V,\quad S^{\vA\vB}(T^\vC)=f^{\vA\vB\vC}{}_\vD T^\vD.
\end{align}
Thus (\ref{pre-GJTS-identity}) 
can be written
\begin{align}
[S^{\vA\vB},\,S^{\vC\vD}]=f^{\vA\vB\vC}{}_\vE S^{\vE\vD}-f^{\vB\vA\vD}{}_\vE S^{\vC\vE}.
\end{align}
For any basis element $T^\vA$, we also define the linear map 
\begin{align}
\bar{T}{}^\vA : V \to \End{V},\quad \bar{T}{}^\vA (T^\vB)=S^{\vA\vB},
\end{align}
Let $L_0$ be the subspace of $\tilde{U}_0$ spanned by all $S^{\vA\vB}$, and let $L_+$ be the subspace of $\tilde{U}_+$ generated by all elements $\bar{T}{}^\vA$ in $\tilde{U}_1$. 
Furthermore,
let $L_-$ be a Lie algebra isomorphic to $L_+$, with the isomorphism denoted by $\tau$.
Thus $L_-$ is generated by all elements $\tau(\bar{T}{}^\vA)$.
Consider the vector space 
\begin{align}
L(V)=L_- \oplus L_0 \oplus L_+. 
\end{align}
We can extend the Lie algebra structures on each of these subspaces to a Lie algebra structure on the whole of $L(V)$, by the relations
\begin{align} \label{sammanfogning}
[S^{\vA\vB},\,\bar{T}{}^\vC]&=f^{\vA\vB\vC}{}_\vD \bar{T}{}^\vD,\nn\\ [\bar{T}{}^{\vA},\,\tau(\bar{T}{}^\vB)]&=S^{\vA\vB},
\nn\\
[S^{\vA\vB},\,\tau(\bar{T}{}^\vC)]&=-f^{\vB\vA\vC}{}_\vD \tau(\bar{T}{}^\vD).
\end{align}
The commutator between two arbitrary elements in two different subspaces $L_+$, $L_-$ or $L_0$ can be derived from (\ref{sammanfogning})
by the Jacobi identity since $L_+$ and $L_-$ are generated by $\bar{T}{}^{\vA}$ and $\tau(\bar{T}{}^\vA)$, respectively.
We can also
extend the isomorphism $\tau$ between the subalgebras $L_-$ and $L_+$ to a graded involution on the Lie algebra $L(V)$. On $L_-$, it is given by the inverse of the original isomorphism, $\tau(\tau(\bar{T}{}^\vA))=\bar{T}{}^\vA$, 
and on $L_0$ by $\tau(S^{\vA\vB})=-S^{\vB\vA}$.

We call $L(V)$ the \textbf{associated Lie algebra} to the generalized triple system $V$. (This definition differs from the 
definition by Kantor of the Lie algebra $\vL(V)$ in \cite{Kantor3.5} in that 
the basis element $\tau(\bar{T}{}^{\vA})$ of $\g_{-1}$
is not the same as the basis element $T^\vA$ of $V$. 
However, if the generalized Jordan triple system $V$ does not contain any
element $a$ such that $(axy)=0$ for all $x,\,y$ in $V$, then we can identify $T^\vA$ with $\tau(\bar{T^\vA})$ and the definitions are equivalent.)

If $V$ is derived from a simple graded Lie algebra $\g$ with a graded involution $\tau$ by (\ref{standardtrippelprodukten}), 
then $L(V)$ is isomorphic to $\g$. Conversely, the \gjts derived from $L(V)$ by (\ref{standardtrippelprodukten}) is isomorphic to $V$ if $V$ is \textbf{simple} in the sense that there are no nontrivial subspace $W\subset V$ such that
$(VVW) \subseteq W$ and $(WVV) \subseteq W$.
This was shown by Kantor in \cite{Kantor3.5} and we also include a proof of the first assertion in the appendix of {\gcrkma} (where we use a somewhat different notation). 
Thus there is a one-to-one correspondence between simple graded Lie algebras
and simple generalized Jordan triple systems.
In the rest of this section we will refine this to a one-to-one correspondence between simple nicely graded Lie algebras and simple normed triple systems.

The basis elements of $L_2$ are
all commutators
$[\bar{T}{}^\vA,\,\bar{T}{}^\vB]$
in the universal graded Lie algebra $\tilde{U}(V)$ of $V$.
Since they are elements in $L_2$, they are linear operators
$V \to L_1$.
This means that 
$[\bar{T}{}^\vA,\,\bar{T}{}^\vB](T^\vC)$ is an element in $L_1$ for any $T^\vC$ and thus
is a linear map $V \to \End{V}$.
We write $[\bar{T}{}^\vA,\,\bar{T}{}^\vB]=\bar{T}{}^{\vA\vB}$.
It follows from the recursively defined commutation relations
for $\tilde{U}(V)$, that the linear map
\begin{align}
\bar{T}{}^{\vA\vB}(T^\vC): V \to \End{V}
\end{align}
is given by
\begin{align}
\bar{T}{}^{\vA\vB}(T^\vC)=
[\bar{T}{}^\vA,\,\bar{T}{}^\vB](T^\vC)&=
(\ad{\bar{T}{}^\vA} \circ \bar{T}{}^{\vB}-
\ad{\bar{T}{}^\vB} \circ \bar{T}{}^{\vA})(T^\vC)\nn\\
&=(\ad{\bar{T}{}^\vA})(\bar{T}{}^{\vB}(T^\vC))
-(\ad{\bar{T}{}^\vB})(\bar{T}{}^{\vA}(T^\vC))\nn\\
&=[{\bar{T}{}^\vA},\,S^{\vB\vC}]
-[{\bar{T}{}^\vB},\,S^{\vA\vC}]\nn\\
&=-f^{\vB\vC\vA}{}_\vD \bar{T}{}^D
+f^{\vA\vC\vB}{}_\vD \bar{T}{}^D \label{gabcd}
\end{align}
where we in the last step have used the commutation relations 
\begin{align}
[S^{\vA\vB},\,\bar{T}{}^\vC]&=f^{\vA\vB\vC}{}_\vD\bar{T}{}^\vD.
\end{align}
We can write this as
\begin{align}
\bar{T}{}^{\vA\vB}(T^\vC)=g^{\vA\vB\vC}{}_\vD \bar{T}{}^\vD,
\end{align}
where $g^{\vA\vB\vC}{}_\vD = f^{\vA\vC\vB}{}_\vD - f^{\vB\vC\vA}{}_\vD$.
We note that the tensor $g^{\vA\vB\vC}{}_\vD$ is antisymmetric in the indices $\vA$ and $\vB$
due to the antisymmetry of the bracket $[T^\vA,\,T^\vB]$.
In 
{\gcrkma} we write (following \cite{Palmkvist:2005gc}) the operator
$\bar{T}{}^{\vA\vB}$ as $\la T^\vA,\,T^\vB \ra$, or generally
\begin{align}
\la u,\,v \ra (x) =(uxv)-(vxu).
\end{align}
If any triple product $(uxv)$ is symmetric in $u$ and $v$ then all operators
$\la u,\,v \ra$ are zero, and the associated Lie algebra $L(V)$ is three-graded,
\begin{align}
L(V)=L_{-1} + L_0 + L_1.
\end{align}
In this special case, $V$ is a \textbf{Jordan triple system} \cite{Jacobson1}. Conversely, if a Lie algebra 
$\g$ is three-graded, then the triple product (\ref{standardtrippelprodukten}) will automatically be symmetric in $x$ and $z$ and thus $\g_{-1}$ is a Jordan triple system.

We return to the general case, where $L_2 \neq 0$.
In the same way as above we can define a basis of $L_3$ consisting of elements
\begin{align}
\bar{T}{}^{\vA\vB\vC}=[\bar{T}{}^{\vA\vB},\,\bar{T}{}^\vC]=[[\bar{T}{}^{\vA},\,\bar{T}{}^{\vB}],\,\bar{T}{}^\vC]
\end{align}
in the universal graded Lie algebra $\tilde{U}(V)$ of $V$, and write
\begin{align}
\bar{T}{}^{\vA\vB\vC}(T^{\vD})(T^{\vE})
=g^{\vA\vB\vC\vD\vE}{}_{\vF} \bar{T}{}^{\vF}
\end{align}
for some tensor $g^{\vA\vB\vC\vD\vE}{}_{\vF}$, which specifies the linear map $\bar{T}{}^{\vA\vB\vC}:V\to \tilde{U}{}_2$ completely.
Again from the antisymmetry of the Lie bracket, we know that $g^{\vA\vB\vC\vD\vE}{}_{\vF}$ must be antisymmetric
in the first two indices. 
Furthermore, the Jacobi identity tells us that $g^{\vA\vB\vC\vD\vE}{}_{\vF}$ vanishes upon antisymmetrization in the first three upper indices. A calculation like (\ref{gabcd}) for $g^{\vA\vB\vC}{}_\vD$ gives
\begin{align}
g^{\vA\vB\vC\vD\vE}{}_{\vF}&=
2(-f^{\vC\vD[\vA}{}_\vG g^{|\vG|\vB]\vE}{}_\vF
+f^{[\vA|\vD|\vB]}{}_\vG g^{\vG\vC\vE}{}_\vF)\nn\\
&=2(-f^{\vC\vD[\vA}{}_\vG f^{|\vG\vE|\vB]}{}_\vF
+f^{[\vA|\vD|\vB]}{}_\vG f^{\vG\vE\vC}{}_\vF\nn\\&\quad\quad
+f^{\vC\vD[\vA}{}_\vG f^{\vB]\vE\vG}{}_\vF
-f^{[\vA|\vD|\vB]}{}_\vG f^{\vC\vE\vG}{}_\vF),
\end{align}
which indeed satisfies $g^{[\vA\vB\vC]\vD\vE}{}_{\vF}=0$.
Continuing in this way, one obtains a recursion formula for the constants
$g^{\vA_1 \cdots \vA_{k} \vB_1 \cdots \vB_{k-1}}{}_{\vB_{k}}$
that appear at each level $k$. It reads
\begin{align}
g^{\vA_1 \cdots \vA_{k}\vB_1 \cdots \vB_{k-1}}{}_{\vB_k}
=g^{\vC\vA_3 \cdots \vA_k \vB_2\cdots\vB_{k-1}}{}_{\vB_k}f^{\vA_1\vB_{1}\vA_2}{}_{\vC}
-\sum g^{\vC_{ij}\vB_2\cdots\vB_{k-1}}{}_{\vB_k} f^{\vA_j\vB_{1}\vA_i}{}_{\vC},
\end{align}
where the sum goes over all $i,\,j$ such that $1 \leq i < j \leq k$
and $\vC_{ij}$ denotes the sequence of indices obtained from $\vA_1
\cdots \vA_{k}$ by omitting $\vA_j$ and replacing $\vA_i$ by $\vC$, that is,
\begin{align}
\vC_{ij}=\vA_1 \cdots \vA_{i-1}\,\vC\,\vC_{i+1} \cdots \vA_{j-1}\,\vA_{j+1} \cdots
\vA_{k}.
\end{align}
We will always have 
\begin{align}
g^{(\vA_1 \vA_2) \vA_3 \vA_4 \cdots \vA_{k}\vB_1 \cdots \vB_{k-1}}{}_{\vB_k}=
g^{[\vA_1 \vA_2 \vA_3] \vA_4 \cdots \vA_{k}\vB_1 \cdots \vB_{k-1}}{}_{\vB_k}=0
\end{align}
due to the antisymmetry of the Lie bracket and the Jacobi identity,
but (for $k\geq4$) the tensor $g^{\vA_1 \cdots \vA_{k}\vB_1 \cdots 
\vB_{k-1}}{}_{\vB_{k}}$ will in addition  satisfy further (anti-)symmetries.
Thus we will in this way for each $k$ obtain a subrepresentation of the tensor product of $k$ vector representations
of $\sl(m)$, where $m=\dim{V}$. The tensor $g$ will be a linear combination of the projectors of the representations that occur at each level.
For example, if $g$ at level $k$ is symmetric under permutation of two of the $k$ first upper indices, say $\vA$ and $\vB$, then
\begin{align}
\bar{T}{}^{\cdots \vA \cdots \vB \cdots }=\bar{T}{}^{\cdots \vB \cdots \vA \cdots }
\end{align} 
at level $k$. We must count these two expressions as one single element at 
level $k$ since they define the same linear map
$V \to \tilde{U}{}_{k-1}$.
On the other hand, they would represent two different elements in the free Lie algebra generated by $\bar{T}^\vA$.
Thus $L_+$ 
is the algebra that we obtain from the free Lie algebra by factoring out the ideal that is recursively defined by $g$ at each level. 

Suppose now that $V$ is a normed triple system with a metric $h$. Since $h$ is non-degenerate, this implies in particular that there is no element $a$ in $V$ 
such that $(axy)=0$ for all $x,\,y$ in $V$, so we can identify $T^\vA$ with $\tau(\bar{T}{}^\vA)$. 
Thus we can write the elements in $L_+$ at level $k$ as $\bar{T}{}^{\vA_1\cdots \vA_k}$ and those at level $-k$ as
${T}{}^{\vA_1\cdots \vA_k}$.
The graded involution is simply given by 
\begin{align}
T^{\vA_1\cdots \vA_k} \leftrightarrow \bar{T}{}^{\vA_1\cdots \vA_k}.
\end{align}
We introduce a bilinear form $\kappa$ on $L(V)$ by
\begin{align} 
\kappa(T^\vA,\,\bar{T}{}^\vB)&=h^{\vA\vB}, &
\kappa(T^{\vA_1\cdots \vA_k},\bar{T}{}^{\vB_1\cdots \vB_k})&=
(-1)^{k+1}g^{\vA_1\cdots \vA_k\vB_k\cdots \vB_{1}},
\end{align}
for any $k\geq2$ (where we have raised the last index with the metric $h$) and $\kappa(L_{m},\,L_{n})=0$ if $m+n\neq0$. Then it follows by the construction of $\kappa$ and the commutation relations in $L(V)$ that $\kappa$ is a symmetric, invariant and non-degenerate bilinear form on $L(V)$, and that it is preserved by the graded involution. We have thus shown that  
any normed triple system gives rise to a nicely graded Lie algebra. Conversely, as we have already seen, any nicely graded Lie algebra gives rise to a normed triple system.


\section{Extensions of generalized Jordan triple systems}\label{extensions}
In 
{\gcrkma}  we define for 
any normed triple system $V$ an infinite sequence of normed triple systems $V^{(n)}$, labeled by a positive integer $n$, 
such that
\begin{align}
\dim{V^{(n)}}=n\,\dim{V}.
\end{align}
Thus we can 
denote the basis elements of $V^{(n)}$ by
$T^{\vA}{}_a$, where $a=1,\,2,\,\ldots,\,n$.
In the special case $n=1$, we can suppress the index $a$ and identify
$V^{(n)}=V^{(1)}$ with $V$.
We introduce a metric $h^{(n)}$ on $V^{(n)}$ by
\begin{align}
h^{\vA}{}_a{}^{\vB}{}_{b} =
h^{(n)}(T^{\vA}{}_{a},\,T^{\vB}{}_{b})=h^{\vA\vB}\delta_{ab}.
\end{align}
The structure constants of $V^{(n)}$ can be expressed in those of $V$ and the metric as
\begin{align}
f^{\vA}{}_a{}^\vB{}_b{}^\vC{}_c{}^{\vD}{}_{d}{}=
f^{\vA\vB\vC\vD}\de_{ab}\de_{cd}
-h^{\vA\vB} h^{\vC\vD} \de_{ab}\de_{cd}
+h^{\vA\vB}h^{\vC\vD}
\de_{bc}\de_{ad}.
\end{align}
One can easily check that $V^{(n)}$ is a normed 
\gjts as well as $V$.
In {\gcrkma} we show that if the Lie algebra associated to $V$ is a finite
Kac-Moody algebra $\mathfrak{h}$, where the grading is given by a simple root 
$\alpha$, then the Lie algebra $\g$ associated to
$V^{(n)}$ is a Kac-Moody algebra as well.
(The subalgebra $\mathfrak{h}$ of $\g$ should not be confused with the Cartan subalgebra, which we called $\mathfrak{h}$ in chapter  \ref{liechapter}.)
Moreover, the Dynkin diagram of $\g$ is obtained from that of $\mathfrak{h}$ 
by adding $n-1$ nodes. 
Each node that we add is connected to the previous one with a single line, starting from the node in the Dynkin diagram of $\mathfrak{h}$ corresponding to the simple root $\alpha$. 
The theorem can easily be generalized to affine and hyperbolic algebras.
In section \ref{e10frome8}, we will apply this method to show how $\frake_{10}$ can be constructed 
from $\frake_{8}$.

\section{Three-algebras and M2-branes}

As we mentioned in section \ref{lagjtssection},
the triple product in a Jordan triple system is symmetric under a permutation of the first and the third element. 
We will not consider such triple systems
here, but instead investigate 
the possibility of a triple product which is \textit{antisymmetric} under a permutation of the the first and the third element,
\begin{align}
(xyz)=-(zyx).
\end{align}
If a generalized Jordan triple system satisfies this identity, then we call it an \textbf{antisymmetric triple system}.
Thus, in addition to the identity (\ref{pre-GJTS-identity}), the structure constants of a normed antisymmetric triple system satisfy
\begin{align} \label{asymstruktur}
f^{\vA\vB\vC\vD}=-f^{\vC\vB\vA\vD}=-f^{\vA\vD\vC\vB}=f^{\vC\vD\vA\vB}
\end{align}
and the identity (\ref{pre-GJTS-identity}) can be written
\begin{align} \label{asymgjtsid}
f^\vE{}_\vD{}^{[\vA}{}_\vC f^{\vB]}{}_\vG{}^{\vD}{}_\vH =
f^{\vA}{}_\vD{}^{\vB}{}_{[\vG} f^{\vE}{}_{\vH]}{}^{\vD}{}_\vC.
\end{align}

Any associative algebra $V$ with an anti-involution $C$ is an antisymmetric triple system under the triple product
\begin{align} \label{kvaterniontrippelprod}
(xyz)=\tfrac12(xC(y)z-zC(y)x).
\end{align}
For example, we can let $C(x)$ be the transpose, inverse or hermitian conjugate of any element $x$ in a matrix algebra $V$, that closes under $C$.
Then it is straightforward to check that the identity (\ref{asymgjtsid}) is satisfied.
An important example is obtained if we take $V$ to be the divison algebra $\H$ of quaternions and $C$ the conjugation, 
which changes sign on the `imaginary units' $i,\,j,\,k$ (but leaves the real numbers unchanged). According to the famous formula
\begin{align}
i^2 = j^2 = k^2 = ijk =-1
\end{align}
we have
\begin{align}
(ijk)
=-(jik)
&= 1,\nn\\
(jk1)
=-(kj1)
&= -i,\nn\\
(k1i)
=-(1ki) 
&= j,\nn\\
(1ij)
=-(i1j)
&= -k
\end{align}
with the triple product (\ref{kvaterniontrippelprod}).
This triple product is not only
antisymmetric in $x$ and $z$,
but also 
in $x$ and $y$ (or $y$ and 
$z$), so it is in fact \textit{totally} antisymmetric.
Moreover, since $\H$ is a \textit{normed} divison algebra there is a positive-definite norm $h$, given by
\begin{align}
h(x,\,y)=\tfrac12(xC(y)+yC(x)),
\end{align}
and it satisfies
\begin{align}
h(w,\,(xyz))=-h(x,\,(wyz)).
\end{align}
This means that 
if we introduce a basis $T^\vA$ and structure constants $f^{\vA\vB\vC\vD}$ as before, where the last index is raised with the metric 
$h^{\vA\vB}=h(T^\vA,\,T^\vB)$, then $f^{\vA\vB\vC\vD}$ is antisymmetric in all four indices.
If we take 
$T^\vA$ to be $i,\,j,\,k,\,1$ for $\vA=1,\,2,\,3,\,4$, then we find that
\begin{align} \label{A4trealgebran}
f^{\vA\vB\vC\vD}&=\ep^{\vA\vB\vC\vD} & h^{\vA\vB}&=\de^{\vA\vB}.
\end{align}
Normed triple systems with totally antisymmetric triple products were recently used by Bagger and Lambert in the construction of a three-dimensional theory, which was proposed to describe multiple M2-branes \cite{Bagger:2006sk,Bagger:2007jr,Bagger:2007vi}. They showed that if the scalar fields take values in such a triple system, called \textbf{three-algebra}, then one can add a non-propagating gauge field such that the resulting theory is maximally supersymmetric. The closure of the supersymmetry algebra was first shown by Gustavsson \cite{Gustavsson:2007vu}, using a different but equivalent algebraic structure. In this approach, the scalar fields and the gauge field take values in two different subspaces, called $\vA$ and $\vB$, respectively, of an algebra $\vA\oplus \vB$. The subspace $\vB$ closes under the product and is a Lie algebra, unlike the full algebra $\vA \oplus \vB$. 
If we consider the three-algebra as a generalized Jordan triple system, then we can in fact identify $\vA$ and $\vB$ with the subspaces $L_{-1}$ and $L_0$, respectively, of the associated Lie algebra. Their direct sum
$L_{-1}\oplus L_0$ 
closes and forms an algebra under a modified Lie bracket, obtained by replacing 
$[x,\,y]$ by 
$[x,\,\tau(y)]$ for all elements $x,\,y$ in $L_{-1}$.

It was later
proven \cite{Nagy,Papadopoulos:2008gh,Gauntlett:2008uf} that the quaternionic triple system (\ref{A4trealgebran}) is the only non-trivial three-algebra with positive definite metric, up to direct sums of such three-algebras. On the other hand, a generalized notion of three-algebras
has gained interest since
Aharony, Bergman, Jafferis and Maldacena (ABJM) \cite{Aharony:2008ug} constructed a superconformal Chern-Simons theory with less supersymmetry, $\vN=6$ instead of the maximal number $\vN=8$ in three dimensions. Bagger and Lambert showed \cite{Bagger:2008se} that this theory can be formulated in terms of a new kind of three-algebras, as a generalization of their original model. 
Such a triple system is what we call a normed antisymmetric triple system in this chapter. 
That is, a triple system whose structure constants satisfy 
(\ref{asymstruktur}) and (\ref{asymgjtsid})
but are not totally antisymmetric.
As we have seen, there are many examples of such triple systems.
In this chapter we have shown that there is a one-to-one-correspondence
between normed triple systems and nicely graded Lie algebras.
We use this correspondence in {\blgjts} to express the ABJM theory (or the three-algebra reformulation by Bagger and Lambert) entirely in terms of the associated graded Lie algebra.

\chapter{The hidden symmetry algebras $\frake_n$} \label{enchapter}

\noindent
Beside the algebras $\mathfrak{a}{}_n\,(n\geq1)$ and $\mathfrak{d}{}_n\,(n\geq4)$, there are three exceptional Lie algebras that are also simple, finite-dimensional and simply laced.
These are called $\frake_6,\,\frake_7$ and $\frake_8$.
Although they do not belong to any infinite class of simple finite-dimensional algebras, they can be viewed as the first three members in an infinite family of Kac-Moody algebras $\frake_n\,(n\geq6)$ with the following Dynkin diagrams.
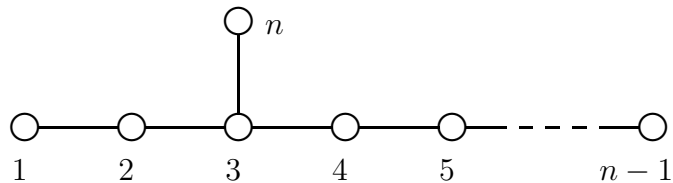
\begin{center}
\scalebox{1}{
\begin{picture}(275,60)
\put(5,-10){$1$}
\put(45,-10){$2$}
\put(85,-10){$3$}
\put(125,-10){$4$}
\put(165,-10){$5$}
\put(225,-10){${n-1}$}
\put(100,45){$n$}
\thicklines
\multiput(10,10)(40,0){5}{\circle{10}}
\multiput(15,10)(40,0){4}{\line(1,0){30}}
\put(175,10){\line(1,0){15}}
\put(230,10){\line(1,0){10}}
\multiput(195,10)(10,0){4}{\line(1,0){5}}
\put(245,10){\circle{10}}
\put(90,50){\circle{10}} \put(90,15){\line(0,1){30}}
\end{picture}}\end{center}
\vspace*{0.4cm}
\noindent
These algebras 
are infinite-dimensional for $n \geq 9$. 
More precisely, $\frake_9$ is affine, $\frake_{10}$ is hyperbolic and 
$\frake_{11}$ is Lorentzian (but not hyperbolic).
In particular 
$\frake_8,\,\frake_9$ and $\frake_{10}$
are interesting from both a mathematical and a physical point of view, and will be devoted one section each of this chapter. First
we will study the general properties of the $\frake_n$ algebras.
These properties are in fact shared by  $\mathfrak{a}{}_4$ and  $\mathfrak{d}{}_5$, if we consider them as
$\frake_4$ and $\frake_5$, respectively. Therefore, when we talk about $\frake_n$ in this chapter, we can assume any 
$n \geq 4$, if nothing else is stated. The case $n=9$ will sometimes be excluded, because $\frake_9$ is affine.
As we saw already in chapter~\ref{hiddenchapter}, the Lie algebras $\frake_{n}$ for $4\leq n \leq 8$ arise as hidden symmetries in the toroidal reduction of maximal supergravity from 11 to $d=11-n$ dimensions. As we will see in the next chapter, there is also evidence for $\frake_{10}$ as a symmetry of the unreduced theory. 

\section{Decomposition under $\mathfrak{a}_{n-1}$}
\label{enundersln}

Since the algebras $\frake_n$ are infinite-dimensional for $n \geq 9$ it is useful to study them as graded algebras where each subspace in the grading is finite-dimensional. We choose the grading given by the node labeled $n$ in the Dynkin diagram, which is also sometimes called the 
\textbf{exceptional node}. This node constitute the difference between $\sl(n)$ and $\frake_n$. Thus, according to chapter \ref{hiddenchapter}, it represents the difference between pure gravity and the bosonic sector of maximal supergravity.
The level decomposition with respect to the exceptional node is crucial for the appearance 
of $\frake_{10}$ in eleven-dimensional supergravity as we will see in chapter
\ref{e10modelchapter}.

Following section \ref{leveldecomp} we consider the level decomposition of the adjoint representation under the subalgebra 
$\g_0{}' = \sl(n)$
corresponding to the horizontal line in the Dynkin diagram.
The subspace $\g_{-1}$ is spanned by all root vectors $e_{\mu}$ such that 
the component of the root $\mu$ corresponding to the
simple
root $\alpha_n$ (the \textbf{exceptional root}) is equal to one.
According to the Chevalley-Serre relations, a basis of $\g_{-1}$ is then the set of all multiple commutators
\begin{align} \label{langbracket}
[[[e_n,\,e_3,\,e_4,\,\ldots,\,e_{i+2}],\,e_2,\,e_3,\,e_4,\,\ldots,\,e_{j+1}],\,e_1,\,e_2,\,e_3,\,e_4,\,\ldots,\,e_{k}]
\end{align}
for all integers $i,\,j,\,k$ such that $0 \leq k \leq j \leq i \leq n-3$
(If $i=0$, this means that the sequence $e_3,\,e_4,\,\ldots,\,e_{i+2}$ should not appear at all, and likewise for $j$ and $k$. For better readability, we have in (\ref{langbracket}) only written out the brackets between these sequences.)
Thus we have
\begin{align}
\dim{\g_{-1}}=\tfrac16{(n-2)(n-1)n}
\end{align}
and since $\g_{-1}$ must be an irreducible representation of $\g_0$, the only
possibility is the totally antisymmetric tensor product of either three vectors or three conjugate vectors.
We write the corresponding tensors as $E^{abc}$ and $F_{abc}$ with the commutation relations
\begin{align}
[K^a{}_b,\,E^{cde}]&=3\de_b{}^{c} E^{ade},&
[K^a{}_b,\,F_{cde}]&=-3\de^a{}_{c} F_{bde}.  \label{enlevel0-1komrel}
\end{align}
Here and throughout this chapter, we use 
\textit{implicit
(anti-)symmerization} which means that the right hand side of an equation is always understood to be (anti-)symmetrized according to the left hand side.
For example, the first equation in (\ref{enlevel0-1komrel}) would otherwise read
\begin{align}
[K^a{}_b,\,E^{cde}]&=3\de_b{}^{[c} E^{|a|de]}
=\de_b{}^{c} E^{ade}+\de_b{}^{d} E^{aec}+\de_b{}^{e} E^{acd}.
\end{align}
Furthermore, the indices will always take the following values,
\begin{align}
a,\,b,\ldots&=1,\,2,\,\ldots,\,n, & i,\,j,\ldots&=1,\,2,\,\ldots,\,n-1.
\end{align}
With our choice
\begin{align}
h_i &= {K^i}_i-{K^{i+1}}_{i+1}, &  e_i &={K^i}_{i+1}, & f_i &={K^{i+1}}_i
\end{align}
(no summation)
for the Chevalley generators of $\sl(n)$,
a solution to the Chevalley relations (\ref{chevalley-rel}) for the remaining Chevalley generators $h_n,\,e_n,\,f_n$ is 
\begin{align}
e_n &= F_{123}, & f_n &=E^{123},&
h_n &= -{K^{1}}_{1}-{K^{2}}_{2}-{K^3}_3+\tfrac{1}{3}K,
\end{align}
where we have embedded $\sl(n)$ in $\gl(n)$ and set
\begin{align}
K={K^1}_1+{K^2}_2+\cdots+{K^n}_n \label{sparet}
\end{align}
for $n \neq 9$. (If $n=9$, we have to embed $\sl(n)$ into a larger algebra, and define $K$ differently, to make $h_n$ linearly independent of the other basis elements in the Cartan subalgebra. We will describe this in section \ref{affinechapter}.)

The Chevalley relation $[e_n,\,f_n]=h_n$ now reads
\begin{align}
[E^{123},\,F_{123}]&={K^{1}}_{1}+{K^{2}}_{2}+{K^3}_3-\tfrac{1}{3}K
\end{align}
and we can covariantize it 
to get an arbitrary $[\g_{1},\,\g_{-1}]$ commutator, 
\begin{align} 
[E^{abc},\,F_{def}]&=18{\delta^{a}}_{d}\de^b{}_e{K^c}_{f}
-2{\delta^{a}}_{d}\de^b{}_e\de^c{}_{f}{K}. \label{[e,f]=h}
\end{align}
Next we want to determine the subspace $\g_{2}$, or the $\sl(n)$ representation $\mathbf{r}{}_2$ that it constitutes. 
We know that it must be contained in the antisymmetric product of
two $\mathbf{r}{}_1$ representations, 
since $\g_{2}$ is spanned by commutators of elements in $\g_{1}$. 
With Dynkin labels for $\sl({10})$, we can write this 
antisymmetric product as
\begin{align} \label{e10dynkinlabeltensprod1}
[(001000000)\times(001000000)]_- &= (000001000)+(010100000).
\end{align}
(It is easy to generalize (\ref{e10dynkinlabeltensprod1}) 
to $n\neq 10$. If $n<10$, we only keep the $n-1$ first indices, and if $n>10$, we add $n-10$ zeros in the end. In addition, the first term on the right hand side vanishes for $n \leq 5$, since it corresponds to a tensor with six antisymmetric indices.)

We can now employ the results in section \ref{lagjtssection} (replacing each $\vA$ index used there with an antisymmetric triple of $\sl(n)$ indices).
If $\frake_n$ is simple (which is the case for $n \neq 9$) then $\frake_n$ is isomorphic to the Lie algebra $L(\g_{-1})$ associated to the triple system $\g_{-1}$ with the triple product
\begin{align}
(E^{abc}E^{def}E^{ghi})=-[[E^{abc},\,F_{def}],\,E^{ghi}].
\end{align}
Thus the elements in $\g_{2}$ are in one-to-one correspondence with the linear maps
$\g_{-1} \to \g_1$ given by
\begin{align}
F_{def} \mapsto [[E^{abc},\,E^{ghi}],\,F_{def}]
\end{align}
for all $E^{abc}$ and $E^{ghi}$ in $\g_{1}$.
Using (\ref{enlevel0-1komrel}), (\ref{[e,f]=h}) and the Jacobi identity,
we have
\begin{align}
[[E^{abc},\,E^{ghi}],\,F_{def}]
&=-54\de^a{}_{d} \de^b{}_e \de^g{}_{f} E^{chi}+6\de^a{}_{d}\de^b{}_e\de^c{}_{f} E^{ghi}\nn\\
&\quad\,+54\de^g{}_{d} \de^h{}_e \de^a{}_{f} E^{ibc}-6\de^g{}_{d}\de^h{}_e\de^i{}_{f} E^{abc}.
\end{align}
It is straightforward to check that the expression on the right hand side 
(antisymmetrized in $[abc]$ and $[ghi]$)
is antisymmetric under permutation of one element from each of the two triples, 
say, $c$ and $g$. Thus it must be antisymmetric in all six upper indices, and $\mathbf{r}_2$ is equal to the first term on the right hand side 
of (\ref{e10dynkinlabeltensprod1}). (This means that level two is empty for $\frake_{4}$ and $\frake_{5}$, and also any higher level.)
Thus we can write
\begin{align}
[E^{abc},\,E^{def}]&=E^{abcdef}=E^{[abcdef]},&
[F_{abc},\,F_{def}]&=-F_{abcdef}=-F_{[abcdef]}.
\end{align}
(The representation $\mathbf{r}{}_{-k}$ is always the conjugate of 
$\mathbf{r}{}_{k}$, and we choose the minus sign since we want $F$ with indices downstairs 
to always be the transpose of $E$ with indices upstairs.)

We proceed to level three. The representation $\mathbf{r}_3$
must be contained in the tensor product $\mathbf{r}_2 \times \mathbf{r}_1$. 
With Dynkin labels for $\sl({10})$
we write this as
\begin{align} \label{e10dynkinlabeltensprod2}
(000001000)\times(000001000) &= (000000001)+(100000010)\nn\\
&\quad\,+(010000100)+(001001000).
\end{align}
Except for $(100000010)$, all 
the irreducible representations on the right hand side are contained in the totally antisymmetric
tensor product of three $\mathbf{r}_1$ representations, and thereby forbidden by the Jacobi identity. Thus $\mathbf{r}_3$ is equal to
the second term in (\ref{e10dynkinlabeltensprod2}). This means that level three (and any higher level) is empty for $\frake_6$ and 
$\frake_7$, since $\mathbf{r}_3$ corresponds to a tensor 
$E^{a|bcdefghi}$ that is antisymmetric in the eight last indices, but vanishes upon antisymmetrization in all nine indices.
We can thus write
\begin{align}
[E^{abc},\,E^{defghi}]&=3E^{[a|bc]defghi}, & 
[F_{abc},\,F_{defghi}]&=-3F_{[a|bc]defghi}, \label{ettan0}
\end{align}
(this normalization will turn out be convenient)
where
$E^{[a|bcdefghi]}=F_{[a|bcdefghi]}=0$, 
or equivalently,
\begin{align}
E^{a|bcdefghi}+8E^{[b|cdefghi]a}=F_{a|bcdefghi}+8F_{[b|cdefghi]a}=0.
\end{align}
Using this, the equations (\ref{ettan0}) can be inverted to
\begin{alignat}{3}
3E^{a|bcdefghi}=4[E^{a[bc},\,E^{defghi]}], &&\quad\quad& \label{ettan02}
3F_{a|bcdefghi}=-4[F_{a[bc},\,F_{defghi]}].
\end{alignat}
We summarize the representation contents (for $n\neq9$) at the first three positive and negative levels:
\begin{alignat}{2}
\ell=3: &\quad E^{a|bcdefghi}&&=E^{a|[bcdefghi]} \nonumber\\
\ell=2: &\quad E^{abcdef}&&=E^{[abcdef]} \nonumber\\
\ell=1: &\quad E^{abc}&&=E^{[abc]} \nonumber\\
\ell=0: &\quad {K^a}_b&&   \nonumber \\                            
\ell=-1: &\quad F_{abc}&&=F_{[abc]} \nonumber\\
\ell=-2: &\quad F_{abcdef}&&=F_{[abcdef]} \nonumber\\
\ell=-3: &\quad F_{a|bcdefghi}&&=F_{a|[bcdefghi]}  \label{bosonfalt}
\end{alignat} 
As we have already mentioned,
some of the generators vanish for $4 \leq n \leq 7$, 
because of the antisymmetries.
For $n=9$, 
there is an additional element at level zero, as we will see in section \ref{affinechapter}.

One could in principle
go on as we have done and determine the representations from the (anti-)symmetries that the basis elements must satisfy. However there are more efficient methods, which are also recursive, but based on information about the roots of the algebra and the weights of the possible representations. In general there is not only one irreducible representation at each level, but a direct sum. For $n \geq 10$, 
the number of representations increases for each level, which soon makes it very complicated to go higher up in levels.
The higher levels of $\frake_{10}$ and 
$\frake_{11}$ have been studied systematically in \cite{West:2002jj,Nicolai:2003fw,Fischbacher:2005fy}.  
Unfortunately, the only pattern that one has been able to find so far is
that
the tensors (\ref{bosonfalt}) generalize to 
\begin{alignat}{2}
\ell=3k+3: &\quad E^{\,\cdots\,|a|bcdefghi}&&=E^{\,\cdots\,|a|[bcdefghi]} \nonumber\\
\ell=3k+2: &\quad E^{\,\cdots\,|abcdef}&&=E^{\,\cdots\,|[abcdef]} \nonumber  \\
\ell=3k+1: &\quad E^{\,\cdots\,|abc}&&=E^{\,\cdots\,|[abc]} \label{gradientrepresentationer}
\end{alignat}
for any $k \geq 0$ and any $n$ (and likewise for the negative levels).
The ellipsis represents $k$ tuples of 9 antisymmetric indices each, and the tensors are symmetric under permutations of the tuples. 
For $n=8$ and $n=9$ there are no other representations, which in particular means that the elements in (\ref{bosonfalt}) for $n=8$ is a basis of $\frake_8$, 
since all 9-tuples vanish.
For $\frake_{10}$, the elements (\ref{gradientrepresentationer}) constitute only a tiny subset
of a basis, since their number grows linearly with the level, while the total number of generators grows exponentially.

We recall from section \ref{chevgensln} that the Killing form for $\sl(n)$, with the choice (\ref{chevgensln-ekv}) of Chevalley generators, 
can be written
\begin{align}
\kappa(K^a{}_b,\,K^c{}_d)&=\de^a{}_d \de^c{}_b + m \de^a{}_b \de^c{}_d 
\end{align}
for an arbitrary constant
$m$. This constant is now fixed by the condition
\begin{align}
2=\kappa(h_n,\,h_n)=\tfrac19(m(n-9)^2+(n+9))
\end{align}
with the solution $m=(9-n)^{-1}$
for $n\neq 9$. (We will come back to the case $n=9$ in section \ref{affinechapter}.)
We also have $\kappa(E^{123},\,F_{123})=1$, which can be covariantized to
\begin{align}
\kappa(E^{abc},\,F_{def})=3! \de^a{}_{d} \de^b{}_e \de^c{}_{f}
\end{align}
By invariance of the Killing form we then get
\begin{align}
\kappa(E^{a_1 \cdots a_6},\,F_{b_1 \cdots b_6})&=6! \, \de^{a_1}{}_{b_1} \cdots \de^{a_6}{}_{b_6},\nn\\
\kappa(E^{a|b_1\cdots b_8},\,F_{c|d_1\cdots d_8})&= \tfrac89 \cdot 8! \,
(\de^a{}_c \de^{b_1}{}_{d_1} \cdots \de^{b_8}{}_{d_8} - \de^{b_1}{}_c \de^{b_2}{}_{d_1} \cdots \de^{b_8}{}_{d_7} \de^a{}_{d_8}).
\end{align}

\section{The maximal compact subalgebra $\mathfrak{k}(\frake_n)$} 
\label{repofken}

We now write $\frake_n$ as the (vector space) direct sum of its maximal compact subalgebra 
$\mathfrak{k}(\frake_n)$ and its orthogonal complement $\mathfrak{p}$ with respect to the Killing form,
\begin{align}
\frake_n = \mathfrak{k}(\frake_n) \oplus \mathfrak{p}.
\end{align}
The projection of an element $2x$ in $\frake_n$ onto $\mathfrak{k}(\frake_n)$ is given by
$x+\omega(x)$, where $\omega$ is the Chevalley involution. Likewise, the projection onto $\mathfrak{p}$ is given by $x-\omega(x)$.

A grading of a Kac-Moody algebra $\g$ with the Chevalley involution $\omega$
as a graded involution leads to a direct sum decomposition of the maximal compact subalgebra
$\mathfrak{k}(\g)$ as well, into subspaces $\mathfrak{k}_0,\,\frakk_1,\,\frakk_2,
\,\ldots$. In this decomposition, $\frakk_k$ is spanned by all elements
$x+\omega(x)\in\g$, where $x \in \g_k$ (and thus $\omega(x) \in \g_{-k}$).
However, we do not call this a grading since the condition 
$[\frakk_m,\,\frakk_n]\subseteq \frakk_{m+n}$ is \textit{not} satisfied. Rather,
\begin{align}  \label{filtrering}
[\frakk_m,\,\frakk_n]\subseteq \frakk_{m+n}+\frakk_{|m-n|}
\end{align}
and likewise for $\mathfrak{p}$.
Nevertheless, we will talk about \textit{levels} of $\mathfrak{k}(\g)$ and $\mathfrak{p}$ referring to the non-negative integers in the decompositions above.
Thus at the first levels 
we have
\begin{align}
J^{ab}&=K^a{}_b-K^b{}_a,\nn\\
J^{abc}&=E^{abc}-F_{abc},\nn\\
J^{abcdef}&=E^{abcdef}-F_{abcdef},\nn\\
J^{a|bcdefghi}&=E^{a|bcdefghi}-F_{a|bcdefghi}
\end{align}
as basis elements for $\mathfrak{k}(\frake_n)$, and
\begin{align}
S^{ab}&=K^a{}_b+K^b{}_a,\nn\\
S^{abc}&=E^{abc}+F_{abc},\nn\\
S^{abcdef}&=E^{abcdef}+F_{abcdef},\nn\\
S^{a|bcdefghi}&=E^{a|bcdefghi}+F_{a|bcdefghi}
\end{align}
for the coset $\mathfrak{p}$. According to (\ref{filtrering}),
the commutator 
\begin{align}
[J^{abc},\,J^{def}]&=J^{abcdef}-18\de^{ad}\de^{be}J^{cf},
\end{align}
of two level one generators is not entirely contained in the level two subspace, but has also a level zero part.

\subsection{Spinor and vector-spinor representations}

The level zero subalgebra of $\frakk(\frake_n)$ is $\so(n)$. It has a 
vector representation
\begin{align} 
J^{ab}v^c=2\de^{ca}v^{b},
\end{align}
and a spinor representation
\begin{align} 
J^{ab}\varphi=\tfrac{1}{2}\Ga^{ab}\varphi,
\end{align}
where $\Ga^{ab}$ is the antisymmetrized product of two $\so(n)$ gamma matrices.
The tensor product of these two representations is the vector-spinor representation 
\begin{align} 
J^{ab}\psi^c=\tfrac{1}{2}\Ga^{ab}\psi+2\de^{ca}\psi^{b}.
\end{align}
In this subsection we will investigate the possibility of extending the spinor and the vector-spinor representation from level zero to the whole of 
$\mathfrak{k}(\frake_n)$. For this it suffices to find an action of the level one generators such that the commutation relations
\begin{align}
[J^{abc},\,J^{def}]&=-18\de^{ad}\de^{be}J^{cf} \label{110villkor}
\end{align}
are satisfied whenever (at least) one of the indices $a,\,b,\,c$ is equal to one of $d,\,e,\,f$. Once we have found such actions of the level one generators, the action of the level two generators is \textit{defined} by the commutation relations
\begin{align}
[J^{abc},\,J^{def}]&=J^{abcdef}
\end{align}
for six distinct indices $a,\,b,\,c,\,d,\,e,\,f$.
Since $\mathfrak{k}(\frake_n)$ is generated by the level one generators, this is enough to define a representation of the whole algebra \cite{Damour:2006xu}.

We write $\Ga^{a_1 a_2 \cdots a_p}=\Ga^{[a_1}\Ga^{a_2} \cdots \Ga^{a_p]}$ for any $p \geq 2$.
It is easy to see that
if we define the action of the level one generator $J^{abc}$ on the 
$\so(n)$ spinor $\varphi$ by
\begin{align}
J^{abc}\varphi =\tfrac12 \Ga^{abc}\varphi,
\end{align}
then the commutation relations of $\frakk(\frake_{n})$ are satisfied.
The first levels of the representation are then \cite{deBuyl:2005zy,Damour:2005zs}
\begin{align}
J^{ab}\varphi &=\tfrac12 \Ga^{ab}\varphi,\nn\\
J^{abc}\varphi &= \tfrac12 \Ga^{abc}\varphi,\nn\\
J^{abcdef}\varphi &= \tfrac12 \Ga^{abcdef}\varphi,\nn\\
J^{a|bcdefghi}\varphi &=4 \de^{ab}\Ga^{cdefghi}\varphi.
\end{align}
The vector-spinor representation is
more complicated.
We make the most general ansatz for $J^{abc}\psi^d$, 
including gamma trace terms, 
\begin{align}
J^{abc}\psi{}^d &= A\Ga^{abc}\psi{}^d+B\de^{da}\Ga^{b}\psi{}^c-C\Ga^{dab}\psi{}^c\nn\\
&\quad 
+E \Ga^{abcd}\Ga^e\psi^e+F\de^{da}\Ga^{bc}\Ga^e\psi^e. \label{apa}
\end{align}
From the condition (\ref{110villkor}) when one of the indices $a,\,b,\,c$ is equal to one of $d,\,e,\,f$, we get the equations
\begin{align}
6E(3D+3A-C)+4FC&=0,\nn\\
6E(3D-3A-2C)+2F(B+2C-3A-3D)&=0,\nn\\
4A^2&=1,\nn\\
6AC-BC-C^2+3EG&=0,\nn\\
B^2+2C^2+2FG&=18,\nn\\
B^2+2BC+24AC+4C^2+4FG&=0,\nn\\
BC+4C^2-6EG&=0, \label{allmaaekv}
\end{align}
where $D=A+(n-3)E-F$ and $G=6A+B+(n-2)C$.
We see that 
independently of $n$, we always have the six solutions
\begin{align}
A&=\pm 1/2, & A&=\pm 1/2, &   A&=\pm 1/2,  \nn\\ 
B&=6,       & B&=-6  ,    &   B&=\pm 4,     \nn\\
F&=\mp 1,   & F&=\pm 3 ,  &   C&=\mp 1,   \nn\\
C&=E=0,     & C&=E=0,      &   E&=F=0. 
\end{align}
We are mostly interested in the last pair of solutions, those without gamma
trace terms.
For the first levels we then get \cite{Damour:2005zs,Buyl:2005mt}
\begin{align} 
J{}^{ab}\psi^{c} &= \tfrac{1}{2}\Gamma^{ab}\psi^c+2\delta^{ca}\psi^{b},\nonumber\\
J{}^{abc}\psi^{d} &= \tfrac{1}{2}\Gamma^{abc}\psi^d+4\delta^{da}\Gamma^{b}\psi^{c}-\Gamma^{dab}\psi^{c},\nonumber\\
J{}^{abcdef}\psi^{g} &= \tfrac{1}{2}\Gamma^{abcdef}\psi^g-10\delta^{ga}\Gamma^{bcde}\psi^{f}+4\Gamma^{gabcde}\psi^{f},\nonumber\\
J{}^{a|bcdefghi}\psi^{j} &=\tfrac{16}{9}
(\Ga^{jbcdefghi}\psi^a-\Ga^{jabcdefgh}\psi^{i})\nn\\
&\quad\,
+4\de^{ab}\Ga^{cdefghi}\psi^j-56\de^{ab}\Ga^{jcdefgh}\psi^{i}\nn\\
&\quad\,
+\tfrac{16}{9}(8\de^{ja}\Ga^{bcdefgh}\psi^{i}
-\de^{jb}\Ga^{cdefghi}\psi^a
+7\de^{jb}\Ga^{acdefgh}\psi^{i}).
\label{vspinor}
\end{align}
We stress that the spinor and the vector-spinor representation are finite-dimensional for all $n$, even for $n\geq9$, when the algebra $\frakk$ itself is infinite-dimensional. (The dimension of the representation is determined by the size of the $\so(n)$ gamma matrices.) This means that the kernel of the representation is nontrivial. But the kernel of a representation is an ideal of the algebra, so we conclude that $\frakk(\frake_n)$ is nonsimple for $n\geq9$, unlike $\mathfrak{k}(\frake_6),\,\mathfrak{k}(\frake_7)$ and 
$\mathfrak{k}(\frake_8)$. 
In {\from} we study the ideals of $\mathfrak{k}(\frake_9)$ corresponding to the spinor and vector-spinor representations.

\section{The finite algebra $\frake_8$} \label{e8section}
Having explained the general properties of the $\frake_n$ algebras, we now start our exposition of $\frake_8$, $\frake_9$ and $\frake_{10}$. First we consider $\frake_8$ with 
the following Dynkin diagram:
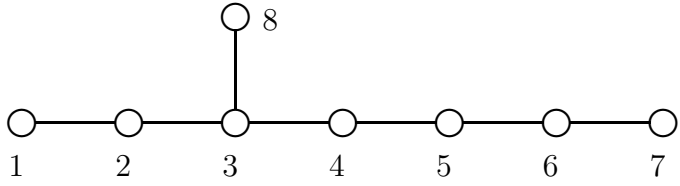
\begin{center}
\scalebox{1}{
\begin{picture}(260,60)
\put(5,-10){$1$}
\put(45,-10){$2$}
\put(85,-10){$3$}
\put(125,-10){$4$}
\put(165,-10){$5$}
\put(205,-10){$6$}
\put(245,-10){$7$}
\put(100,45){$8$}
\thicklines
\multiput(10,10)(40,0){7}{\circle{10}}
\multiput(15,10)(40,0){6}{\line(1,0){30}}
\put(90,50){\circle{10}} \put(90,15){\line(0,1){30}}
\end{picture}}\end{center}
\vspace*{0.4cm}
As we have already mentioned, $\frake_8$ is the largest of the finite-dimensional exceptional Lie algebras.
Another characteristic property of $\frake_8$ is its self-dual root lattice. This fact is crucial for the anomaly cancellation in the heterotic $E_8 \times E_8$ string theory. However, it also makes $\frake_8$ calculations very complicated, since the smallest non-trivial irreducible representation of $\frake_8$ is its adjoint representation, which is 248-dimensional.
 
We let $t^\vA$ be a basis of $\frake_8$, for $\vA=1,\,2,\,\ldots,\,248$. We denote the structure constants and the components of the Killing form in this basis by $f$ and $\eta$, respectively,
\begin{align}
[t^\vA,\,t^\vB]&=f^{\vA\vB}{}_\vC t^\vC & \kappa(t^\vA,\,t^\vB)&=\eta^{\vA\vB}.
\end{align}
We use $\eta^{\vA\vB}$ to raise adjoint $\frake_8$ indices and its inverse $\eta_{\vA\vB}$ to lower them.
For example, $f^{\vA\vB\vC}=f^{\vA\vB}{}_\vD \eta^{\vC\vD}$. It follows from the invariance of the Killing form that $f$ with all indices upstairs (or all indices downstairs) is totally antisymmetric, that is, antisymmetric in any two indices that are both upstairs or both downstairs, in particular
$f^{\vA\vB\vC}=f^{[\vA\vB\vC]}$.

The tensors $f$ and $\eta$ are examples of \textit{invariant tensors} of the adjoint representation. Consider a tensor $S^{\vA\vB}$ with two adjoint $\frake_8$ indices, that is, the tensor product of two adjoint $\frake_8$ representations. We have
\begin{align}
t^\vA(S^{\vB\vC})=-2f^{\vA\vB}{}_\vD S^{\vC\vD}
\end{align}
and it follows by the Jacobi identity and the invariance of the Killing form that
\begin{align}
t^\vA(\eta_{\vB\vC}S^{\vB\vC})=0.
\end{align}
Likewise, for the tensor product of three adjoint representations we have
\begin{align}
t^\vA(f_{\vB\vC\vD}S^{\vB\vC\vD})=0.
\end{align}
Thus $\eta_{\vB\vC}S^{\vB\vC}$ and $f_{\vB\vC\vD}S^{\vB\vC\vD}$ transform as singlets, in the trivial one-dimensional representation.

The tensor product of adjoint $\frake_8$ representations can be decomposed into a direct sum of irreducible representations.
For each irreducible representation there is a projector $\mathbb{P}$, 
such that $\mathbb{P}\mathbb{P}=\mathbb{P}$, whereas $\mathbb{P}\mathbb{Q}=0$ if $\mathbb{P}$ and $\mathbb{Q}$ are projectors that correspond to two different (but possibly equivalent) irreducible representations. 
For the tensor product of two adjoint representations, the projector of the singlet has the components
\begin{align}
(\mathbb{P}_{\bf1})_{\vA\vB}{}^{\vC\vD} &= \tfrac{1}{248}\eta_{\vA\vB}\eta^{\vC\vD}.
\end{align}
The full decomposition of the tensor product reads
\begin{align}
\bf248 \times \bf248 = \bf1 + \bf248 + \bf3875 +\bf27000 + \bf30380
\end{align}
and the corresponding projectors have the components 
\begin{align} \label{e8projektorer}
(\mathbb{P}_{\bf1})_{\vA\vB}{}^{\vC\vD} &= \tfrac{1}{248}\eta_{\vA\vB}\eta^{\vC\vD},\nn\\
(\mathbb{P}_{\bf248})_{\vA\vB}{}^{\vC\vD} &= -\tfrac{1}{60}
f^{\vE}{}_{\vA\vB}f_{\vE}{}^{\vC\vD},\nn\\
(\mathbb{P}_{\bf3875})_{\vA\vB}{}^{\vC\vD} &= \tfrac{1}{7}\delta_{(\vA}{}^{\vC}\delta_{\vB)}{}^{\vD}
-\tfrac{1}{56}\eta_{\vA\vB}\eta^{\vC\vD}-\tfrac{1}{14}f^{\vE}{}_{\vA}{}^{(\vC}f_{\vE\vB}{}^{\vD)},\nn\\
(\mathbb{P}_{\bf27000})_{\vA\vB}{}^{\vC\vD} &= \tfrac{6}{7}\delta_{(\vA}{}^{\vC}\delta_{\vB)}{}^{\vD}
+\tfrac{3}{217}\eta_{\vA\vB}\eta^{\vC\vD}+\tfrac{1}{14}f^{\vE}{}_{\vA}{}^{(\vC}f_{\vE\vB}{}^{\vD)},\nn\\
(\mathbb{P}_{\bf30380})_{\vA\vB}{}^{\vC\vD} &= 
\delta_{[\vA}{}^{\vC}\delta_{\vB]}{}^{\vD}+\tfrac{1}{60}f^{\vE}{}_{\vA\vB}f_{\vE}{}^{\vC\vD},
\end{align}
which were given in \cite{Koepsell:1999uj}.
If we consider the symmetric product of four or six adjoint representations then we will also find a singlet, and the corresponding projector will also be possible to write entirely in terms of the invariant tensor $\eta$, analogously to the one above. However, in the symmetric product of eight adjoint 
$\frake_8$ representations, there will be an additional singlet, which is not possible to express in $\eta$. We must introduce a new invariant tensor, which has eight symmetric indices. In {\octic} we give an explicit expression for this \textit{primitive} invariant of order eight.

\subsection{Decomposition under $\mathfrak{a}{}_7$ and $\so(16)$}
According to 
section \ref{enundersln}, we have the following basis of $\frake_8$ in the $\sl(8)$ decomposition:
\begin{align} 
\ell=3: &\quad Z^{a|bcdefghi}, \nonumber\\
\ell=2: &\quad Z^{abcdef},\nn\\
\ell=1: &\quad Z^{abc},\nn\\
\ell=0: &\quad {G^a}_b,\nn\\
\ell=-1: &\quad Z_{abc},\nn\\
\ell=-2: &\quad Z_{abcdef},\nn\\
\ell=-3: &\quad Z_{a|bcdefghi}.
\end{align} 
The generators at level $\ell=\pm 3$ can be dualized to
\begin{align}
Z^a&=\tfrac1{8!}\ep_{bcdefghi}Z^{a|bcdefghi}, & Z_a&=\tfrac1{8!}\ep^{bcdefghi}Z_{a|bcdefghi}.
\end{align}
(Instead of $E,\,F,\,K$ we here use the notation $Z,\,G$ as in \cite{Koepsell:1999uj}. This will be convenient in the next chapter when we compare 
$\frake_8$ to $\frake_9$.)
The basis of $\frakk(\frake_8)$ is thus
\begin{align}
J^{ab}&={G^a}_b-{G^b}_a,\nn\\
J^{abc}&=Z^{abc}-Z_{abc},\nn\\
J^{abcdef}&=Z^{abcdef}-Z_{abcdef},\nn\\
J^{a}&=Z^{a}-Z_{a}
\end{align}
but we could as well replace the single indices on the level three generator by seven antisymmetric indices. Then we would have tensors with $2,\,3,\,6$ or $7$ antisymmetric $\sl(8)$ indices. This is reminiscent of the
Clifford algebra generated by eight elements, which can be (faithfully) represented by all $16 \times 16$ matrices. The eight generators $\Ga^a$ are symmetric matrices, which anticommute and square to one.
This means that the antisymmetrized products 
\begin{align}
\Ga^{a_1 a_2 \cdots a_p}=\Ga^{[a_1}\Ga^{a_2} \cdots \Ga^{a_p]}
\end{align}
with $p=2,\,3,\,6$ or $7$ antisymmetric indices constitute a basis of the subspace of all antisymmetric $16 \times 16$ matrices. This subspace closes under the commutator and form the Lie algebra $\so(16)$. It is therefore natural to guess that the maximal compact subalgebra $\frakk(\frake_8)$ is isomorphic to
$\so(16)$. If we define a new basis by
\begin{align} \label{e8basrelation}
4X^{IJ}=-\tfrac1{2!}\Ga^{ab}{}_{IJ}J^{ab}
-\tfrac1{3!}\Ga^{abc}{}_{IJ}J^{abc}
-\tfrac1{6!}\Ga^{abcdef}{}_{IJ}J^{abcdef}
-(\Ga^a\Ga^{9}){}_{IJ}J^{a},
\end{align}
where $X^{IJ}=-X^{JI}$ and
\begin{align} 
\Ga^9=\Ga^1\Ga^2\cdots \Ga^8=
\begin{pmatrix}1&0\\0&-1\end{pmatrix},
\end{align}
then we get
\begin{align}
[X^{IJ},\,X^{KL}] &= 4 \delta^{JK}X^{IL},
\end{align}
which are indeed the commutation relations of $\so(16)$.
The relation (\ref{e8basrelation}) can be inverted to give
\begin{align} 
J^{ab}&=\tfrac{1}{4}{\Gamma^{ab}}_{IJ}X^{IJ},\nonumber\\\nonumber
J^{abc}&=-\tfrac{1}{4}{\Gamma^{abc}}_{IJ}X^{IJ},\\\nonumber
J^{abcdef}&=\tfrac{1}{4}{\Gamma^{abcdef}}_{IJ}X^{IJ},\\
J^{a}&=-\tfrac{1}{4}({\Gamma^{a}\Gamma^{9}})_{IJ}X^{IJ}. \label{komb1}
\end{align}
The matrices $\Ga^a$ can be decomposed into $\so(8)$ gamma matrices as
\begin{align} \label{so9gamma}
\Ga^a{}_{IJ} =
\begin{pmatrix}
0&\ga^a{}_{\al\dot{\al}}\\\ga^a{}_{\da\al}&0
\end{pmatrix},
\end{align}
where $\ga^a{}_{\da\beta}$ is the transpose of $\ga^a{}_{\al\dot{\beta}}$, and $\al,\,\dot{\al}=1,\,2,\,\ldots,\,8$.
Then the relations (\ref{komb1}) can be written in an $\so(8)$ covariant form as
\cite{Koepsell}
\begin{align}
J^{ab} &= \tfrac14 \ga^{ab}{}_{\al\beta} X^{\al\beta}+\tfrac14 \ga^{ab}{}_{\da\dot{\beta}} X^{\da\dot{\beta}},\nn\\
J^{abc} &= -\tfrac12 \ga^{abc}{}_{\al\dot{\beta}} X^{\al\dot{\beta}},\nn\\
J^{abcdef} &= 
\tfrac14 \ga^{abcdef}{}_{\al\beta} X^{\al\beta} + \tfrac14 \ga^{abcdef}{}_{\da\dot{\beta}} X^{\da\dot{\beta}},\nn\\
J^a &= -\tfrac12  (\ga^a\ga^9){}_{\al\dot{\beta}} X^{\al\dot{\beta}}. \label{JX}
\end{align}
Due to $\so(8)$ triality, the matrices $\ga^\al{}_{a \dot{\al}}$ and $\ga^{\dot{\al}}{}_{a \al}$ have the same properties as $\ga^a{}_{\da\al}$. 
Thus 
we can take as $\so(16)$ gamma matrices the tensor products
\begin{align}
\hat{\Ga}{}^{\alpha} &= \delta \otimes \ga^\al,&
\hat{\Ga}{}^{\da} &= \ga^\da \otimes \ga^9,
\end{align}
(where $\delta$ is the $8 \times 8$ identity matrix)
with the components
\begin{align} \label{so16gamma}
\hat{\Ga}{}^\al{}_{\beta\dot{\al}|\ga a} &= \delta_{\beta\gamma} \ga^a{}_{\al\dot{\al}},
 & 
\hat{\Ga}{}^{\da}{}_{ab|\al c} & = \delta_{bc} \ga^a{}_{\al\dot{\al}},\nn\\
\hat{\Ga}{}^\al{}_{ba|c\dot{\al}} &= \delta_{bc} \ga^a{}_{\al\dot{\al}},
 & \hat{\Ga}{}^{\da}{}_{\al\dot{\be}|a\dot{\gamma}} &= - \delta_{\dot{\be}\dot{\ga}} \ga^a{}_{\al\dot{\al}},
\end{align}
as in \cite{Nicolai:1986jk}, and all other components are zero.
The splits of the $\so(16)$ vector, spinor and cospinor indices $(I,\,A,\,\dot{A})$
into $\so(8)$ indices,
\begin{align}
I &\to (\al,\,\dot{\al}),&
A &\to (\al\dot{\al},\,ab),&
\dot{A} &\to (\al a,\, b \dot{\al}),
\end{align}
are in accordance with the decompositions
\begin{align}
{\bf 16} &\rightarrow  {\bf {8}_s} \oplus {\bf {8}_c},\nn\\
{\bf 128_s} &\rightarrow {\bf (8_v \otimes {8}_v)} \oplus {\bf (8_v \otimes {8}_v)},\nn\\
{\bf 128_c} &\rightarrow {\bf (8_v \otimes {8}_s)} \oplus {\bf (8_c \otimes {8}_v)}
\end{align}
of these $\so(16)$ representations under the diagonal $\so(8) \oplus \so(8)$ subalgebra.
If we now introduce a new basis of the coset given by the relations
\begin{align} 
Y^{ab} &= S^{ab} -\tfrac16 \de^{ab} S^{kk} +\tfrac1{1440}
\ep^{abcdefgh} S^{cdefgh},\nn\\
Y^{\al\db} &= \tfrac14 \ga^a{}_{\al\db} S^a -\tfrac1{24}
\ga^{abc}{}_{\al\db}S^{abc}, 
\end{align}
which can be inverted to \cite{Koepsell}
\begin{align}
S^{ab} &= 2 Y^{(ab)} - \delta^{ab} Y^{cc},\nn\\
S^{abc} &= -
  \tfrac12\ga^{abc}{}_{\al\dot{\beta}}Y^{\al\dot{\beta}},\nn\\
S^{abcdef} &=  \ep^{abcdefgh}
  Y^{gh},\nn\\
S^a &=-\tfrac12 \ga^a{}_{\al\dot{\beta}} Y^{\al\dot{\beta}}, \label{SY}
\end{align}
then we find that the $\frake_8$ commutation relations become
\begin{align}\label{E8decom}
  [X^{IJ},\,X^{KL}]  &= 4 \delta^{JK}X^{IL},
 \\ \nonumber
  [X^{IJ},\,Y^{A}] &= -\tfrac{1}{2}\hat{\Gamma}^{IJ}{}_{AB}Y^{B},  \\
  \nonumber
  [Y^{A},\,Y^{B}] &= \tfrac{1}{4}\hat{\Gamma}^{IJ}{}_{AB}X^{IJ}.
\end{align}
(Note that we use a different notation in {\octic}.)
Thus the adjoint $E_8$ representation decomposes as
\begin{align}
\bf248 \rightarrow \bf120+128_s,
\end{align}
under the maximal compact subalgebra $\frakk(\frake_8)=\so(16)$ (but we could of course have chosen ${\bf128_c}$ as well as ${\bf128_s}$).
Accordingly, we split the adjoint $E_8$ indices as
\begin{align}
\vA \to ([IJ],\,A).
\end{align}
When the indices appear in a summation, we must also include a factor of $1/2$ for each antisymmetric pair $[IJ]$, to avoid double-counting. Furthermore, in any equation where such index pairs appear on the left hand side, we will
follow our convention of
implicit
antisymmetrization on the right hand side.
Thus the structure constants can be written
\begin{align}
  f^{IJ\,KL}{}_{MN} &=  8\delta^{IK}\delta^{LM}\delta^{JN},
  &
  f^{IJ\,A}{}_{B}& = -\tfrac{1}{2}\Gamma^{IJ}{}_{AB},
\end{align}
and the components of the Killing form become
\begin{align}\label{Cartan-Killing}
  \eta^{AB}&=\delta^{AB}, &
  \eta^{IJ\,KL}&=-2\delta^{IK}\delta^{JL}.
\end{align}
This means that we have to change sign when we raise or lower an antisymmetric pair $[IJ]$ of $\so(16)$ vector indices with the $\frake_8$ Killing form. (For the spinor indices, upstairs or downstairs does not matter.)

Before proceeding to $\frake_9$,
we return to the $\frake_8$ basis that we obtain from the level decomposition under $\sl(8)$. In this basis, the $\so(16)$ vector and 
spinor representations
\begin{align} \label{so16rep}
X^{IJ}\varphi^K &= -2 \delta^{JK}\varphi^I,\nn\\
X^{IJ}\chi^{\dot{A}} &= \tfrac{1}{2}\Ga^{IJ}{}_{\dot{A}\dot{B}}\chi^{\dot{B}}
\end{align}
are given by
\begin{align}
J^{ab}\varphi &= \tfrac{1}{2}\Ga^{ab}\varphi,\nn\\
J^{abc}\varphi &= -\tfrac{1}{2}\Ga^{abc}\varphi,\label{nlikamedattas}
\end{align}
\begin{align}
J{}^{ab}\chi^{c} &= \tfrac{1}{2}\Gamma^{ab}\chi^c+2\delta^{c[a}\chi^{b]},\nn\\
J^{abc}\chi^d &= -\tfrac{1}{4}\Ga^{e}\Ga^{abc}\Ga^{d}\chi^e
\nn\\
&= \tfrac{1}{2}\Ga^{abc}\chi^l+3\delta^{d[a}\Ga^{b}\chi^{c]}
-\tfrac{3}{2}\Ga^{d[ab}\chi^{c]}\nn\\
&\quad -\tfrac{1}{4}\Ga^{abcd}\Ga^e\chi^e 
-\tfrac{3}{4}\delta^{d[a}\Ga^{bc]}\Ga^e\chi^e \label{nlikamedattavs}
\end{align}
at the first levels.
We see that (\ref{nlikamedattavs}) indeed is a solution to (\ref{allmaaekv}) in the case $n=8$.
We can also see that (\ref{nlikamedattavs}) is equivalent to (\ref{vspinor}) 
for $n=8$ by setting
$\chi^a = 2\psi^a-\Ga^a \Ga^b \psi^b$.

\section{The affine algebra $\frake_9$} \label{affinechapter}

In this section, we will study the affine Kac-Moody algebra $\frake_9$, which has the following Dynkin diagram.
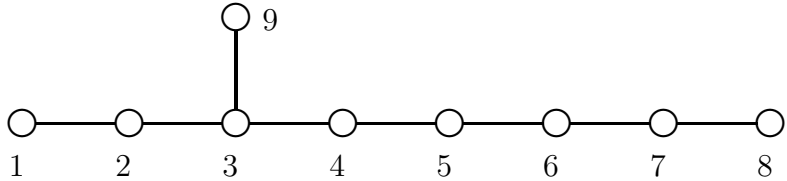
\begin{center}
\scalebox{1}{
\begin{picture}(390,60)
\put(45,-10){$1$}
\put(85,-10){$2$}
\put(125,-10){$3$}
\put(165,-10){$4$}
\put(205,-10){$5$}
\put(245,-10){$6$}
\put(285,-10){$7$}
\put(325,-10){$8$}
\put(140,45){$9$}
\thicklines
\multiput(50,10)(40,0){8}{\circle{10}}
\multiput(55,10)(40,0){7}{\line(1,0){30}}
\put(130,50){\circle{10}} \put(130,15){\line(0,1){30}}
\end{picture}}\end{center}
\vspace*{0.4cm}
However, the algebra constructed from this Dynkin diagram (or equivalently the corresponding Cartan matrix) via the Chevalley-Serre relations as we described in section \ref{chevser} 
is
\textit{not} $\frake_9$, but the \textit{derived} Kac-Moody algebra ${\frake}{}_9{}'$ \cite{Kac}. We will first study ${\frake}{}_9{}'$, and then define 
$\frake_9$ as an extension of ${\frake}{}_9{}'$.

\subsection{Decomposition under $\mathfrak{e}_8$}

We consider the grading 
of ${\frake}{}_9{}'$ given by 
the root corresponding to the rightmost node in the Dynkin diagram above. If we delete this node, we obtain the Dynkin diagram of $\frake_8$,
which means that
the subalgebra $({\frake}{}_9{}')_0$ is the direct sum of $\frake_8$ and a one-dimensional subalgebra spanned by an element in the Cartan subalgebra. 

Counting the basis elements of $({\frake}{}_9{}')_1$, we find that it is 248-dimensional, thus the representation $\mathbf{r}_1$ of $\frake_8$ is (since it is irreducible) the 
adjoint representation. In fact, there is a graded involution $\tau$ 
on $\frake_9{}'$ (which is not the Chevalley involution), such that the subspace $(\frake_9{}')_{-1}$ with the triple product
\begin{align}
(uvw)=[[u,\,\tau(v)],\,w]
\end{align}
is a triple system isomorphic to $\frake_8$ with the triple product
\begin{align}
(xyz)=[[x,\,y],\,z].
\end{align}
Moreover, we find that the bilinear form on $(\frake_9{}')_{-1}$ associated to 
this graded involution $\tau$, defined in (\ref{bilformen}),
is the Killing form on
$\frake_8$.
Thus we can write the structure constants of the triple system as
\begin{align} 
f^{\vA\vB\vC}{}_\vD = f^{\vA\vB}{}_\vE f^{\vE\vC}{}_\vD,
\end{align}
where $f$ with three indices are the $\frake_8$ structure constants introduced in the preceding section,
and we can raise and lower the indices with the Killing form 
$\eta$ on $\frake_8$.
Following the discussion in chapter \ref{gjtschapter} the
$\frake_8$ representation ${\bf r}_2$
at level two is given by the tensor
\begin{align}
g^{\vA\vB\vC}{}_\vD = 2f^{[\vA|\vC}{}_\vE f^{\vE|\vB]}{}_\vD = f^{\vA\vB}{}_\vE f^{\vE\vC}{}_\vD,
\end{align}
where we have used the Jacobi identity.
Thus we have
\begin{align}
g^{\vA\vB}{}_{\vC\vD}=-60 (\mathbb{P}{}_{\bf248})^{\vA\vB}{}_{\vC\vD},
\end{align}
where $\mathbb{P}{}_{\bf248}$ projects 
out the adjoint representation ${\bf 248}$
from the tensor product 
of two such representations
(see the preceding section).
This means that not only $\mathbf{r}_1$ at level one, but also $\mathbf{r}_2$ at level two, is the adjoint representation 
$\mathbf{248}$
of $\frake_8$. If we continue, we will find that we have infinitely many levels and $\mathbf{r}_p=\mathbf{248}$ for any nonzero 
level $p$. We will also find that there is a vector space isomorphism 
$\varphi_p : \frake_8 \to ({\frake}{}_9{}')_p$ for each level $p \neq 0$, 
and for $p=0$ an injective homomorphism 
such that 
\begin{align} \label{e8e9isom}
[x_{(p)},\,y_{(q)}] &= [x,\,y]_{(p+q)} + p\de_{p+q}\kappa(x,\,y)c,
\end{align}
where $x_{(p)}=\varphi_p(x)$ for any $\frake_8$ element $x$ and $c$ is 
the element in the Cartan subalgebra that commutes with $\frake_8$ at level zero. This Cartan element has the form
\begin{align} \label{dencentralalinkomben}
2h_1+4h_2+6h_3+5h_4+4h_5+3h_6+2h_7+h_8+3h_9.
\end{align}
But this element commutes not only with the $\frake_8$ subalgebra, but also with 
$e_2$ (and $f_2$), the simple root vector that defines the grading, and thus with the whole algebra. In other words, it is a \textit{central} element. The existence of a central element in the algebra is due to the vanishing determinant of the Cartan marix. 
It means in particular that the algebra is not simple since a central subspace is also an ideal.
Furthermore, it means that the Killing form is degenerate (since it is invariant).
We can extend the algebra by adding an extra basis element $d$ to the Cartan subalgebra, such 
that $\kappa(c,\,d)=1$ and $\kappa(x,\,d)=0$ for all non-central elements $x$.
Then the Killing form will be
non-degenerate again on the extended algebra. 
It follows by the
invariance of the Killing form that 
the affine root vectors $e_8$ and $f_8$ must be eigenvectors of the adjoint map of $d$, with the eigenvalues $1$ and $-1$, respectively,
\begin{align}
[d,\,e_8]&=e_8,& [d,\,f_8]&=-f_8 \label{derivation}.
\end{align}
Furthermore, we must have 
$[d,\,x]=0$ for all elements $x$ that are orthogonal to $e_8$ and $f_8$. 
Thus $d$ will be an element in the Cartan subalgebra and a characteristic element with respect to the grading given by $\alpha_8$ (which otherwise would be lacking).
The extension of the derived Kac-Moody algebra ${\frake}{}_9{}'$ that we obtain in this way is the affine Kac-moody algebra $\frake_9$. The element $d$ that we add is called \textbf{derivation}. Conversely, the derived algebra can be defined by 
${\frake}{}_9{}'=[\frake_9,\,\frake_9]$ since $d$ never appears on `the right hand side' of any commutation relations.
Using the isomorphism between $\frake_8$ and each subspace in the grading given by $\alpha_8$, 
the commutation relations read
\begin{align} \label{e8e9isom2}
[x_{(p)},\,y_{(q)}] &= [x,\,y]_{(p+q)} + p\de_{p+q}\kappa(x,\,y)c, & [d,\,x_{(p)}]&=p x_{(p)}. 
\end{align}
The Killing form is given by
\begin{align} \label{stromkilling}
\kappa(x_{(p)},\,y_{(q)})&=\de_{p+q}\kappa(x,\,y), & \kappa(c,\,d)&=1
\end{align}
and otherwise zero.
Although the Killing form on $\frake_9$ now is non-degenerate, the subspace spanned by $c$ is still an ideal. The easiest way to modify 
${\frake}{}_9{}'$, such that we get a simple Lie algebra, is to factor out the central subspace. The resulting Lie algebra is called 
the \textbf{current algebra} or \textbf{loop algebra} of $\frake_8$, and 
the commutation relations reduce to $[x_{(p)},\,y_{(q)}] = [x,\,y]_{(p+q)}$.

\subsection{Decomposition under $\mathfrak{a}_8$}

As we mentioned already in section \ref{enundersln}, in the case $n=9$ we cannot define the Cartan elements as
\begin{align} \label{chevgensln-ekv2}
h_i &= {K^{i+1}}_{i+1}-{K^i}_i, & h_n &= -{K^{1}}_{1}-{K^{2}}_{2}-{K^3}_3+\tfrac{1}{3}K,
\end{align}
(no summation), as we did for $n\neq9$. The reason for this is that $-2h_9$ then would be the linear combination
(\ref{dencentralalinkomben})
of the other basis elements in the Cartan subalgebra. 
Thus we get only the current algebra of $\frake_8$ by this construction, and not the full $\frake_9$. To obtain $\frake_9$, we have to add the 
central element $c$ by hand. If we then set
\begin{align}
K=K^a{}_a + c = K^1{}_1 +K^2{}_2 +\cdots+K^9{}_9 + c,
\end{align}
we can keep all expressions that we had for $n \neq 9$, and the resulting algebra is $\frake_9$.
One natural way to do this is to embed $\sl(9)$ in $\sl(10)$ and set
$c = K^{10}{}_{10}$.
Furthermore, we easily see that the derivation must be given by
$d = -K^{9}{}_{9}$.
Then, with the normalization
$\kappa(K^a{}_b,\,K^c{}_d)=\de^c{}_b\de^a{}_d - \de^a{}_b\de^c{}_d$
we get 
$\kappa (c,\,c) = \kappa (d,\,d) =0$ and  
$\kappa (c,\,d)=1$
as we should. 

We recall from section \ref{enundersln} that in the $\sl(n)$ decomposition of $\frake_n$ (for any $n$), the following representations appear at level $\ell$ for any $k\geq0$,
\begin{alignat}{2}
\ell=3k+3: &\quad E^{\,\cdots\,|a|bcdefghi}&&=E^{\,\cdots\,|a|[bcdefghi]}, \nonumber\\
\ell=3k+2: &\quad E^{\,\cdots\,|abcdef}&&=E^{\,\cdots\,|[abcdef]}, \nonumber  \\
\ell=3k+1: &\quad E^{\,\cdots\,|abc}&&=E^{\,\cdots\,|[abc]}. \label{gradientrepresentationer2}
\end{alignat}
The ellipsis represents $k$ antisymmetric 9-tuples of indices. In the case $n=9$ we can thus fix one ordering of the indices in each 9-tuple and then we only need to specify the number $k$ of 9-tuples, or the $\sl(n)$ level $\ell$. Furthermore, we can dualize the $\sl(9)$ tensors with the epsilon tensor at the first positive levels and write
\begin{align}
&E_{abcdef}=\tfrac{1}{3!}\varepsilon_{abcdefghi}E^{ghi} &&\Leftrightarrow&& E^{abc}=\tfrac{1}{6!}\varepsilon^{abcdefghi}E_{defghi},\nonumber\\
&E_{abc}=\tfrac{1}{6!}\varepsilon_{abcdefghi}E^{defghi} &&\Leftrightarrow&& E^{abcdef}=\tfrac{1}{3!}\varepsilon^{abcdefghi}E_{ghi},\nonumber\\
&{E^a}_b=\tfrac{1}{8!}\varepsilon_{bcdefghij}E^{a|cdefghij} &&\Leftrightarrow& &E^{a|bcdefghi}=\varepsilon^{bcdefghij}{E^a}_{j} \label{generatorer}
\end{align} 
(and likewise for the negative levels). Then we see that we actually have the same representations at level $k$ and level 
$k+3$ for any integer $k$ (also $k<0$), except for the two singlets at level zero. Therefore, we can use the letter $E$ for generators at negative as well as positive levels, and write the $\sl(n)$ level $\ell$ as an extra superscript within parentheses.
Then, with a suitable normalization of the generators 
we have
\begin{alignat}{3}
[E^{(3p+1)}{}^{abc},\,E^{(3q-1)}{}_{def}]&=18\de^a{}_d\de^b{}_e E^{(3p+3q)}{}^c{}_f+6p\de_{p+q}\de^a{}_d\de^b{}_e \de^c{}_f c,\nn\\
[{{E^{(3p)}{}^a}_{b}},\,E^{(3q+1)}{}^{cde}]&=3{\delta^c}_bE^{(3p+3q+1)}{}^{ade}, 
\nn\\
[{{E^{(3p)}{}^a}_{b}},\,E^{(3q-1)}{}_{cde}]&=-3{\delta^a}_cE^{(3p+3q-1)}{}_{bde},\nn\\
[{{E^{(3p)}{}^a}_{b}},\,{{E^{(3q)}{}^c}_{d}}]&=\de^c{}_bE{}^{(3p+3q)}{}^a{}_d - \de^a{}_dE^{(3p+3q)}{}^c{}_b,\nn\\
[E^{(3p+1)}{}^{abc},\,E^{(3q+1)}{}^{def}]&=E^{(3p+3q+2)}{}^{abcdef}, 
\nn\\
[E^{(3p-1)}{}_{abc},\,E^{(3q-1)}{}_{def}]&=-E^{(3p+3q-2)}{}_{abcdef}
\end{alignat}
for any integers $p,\,q$. Up to dualization, and toghether with
$[c,\,x]=0$ for any $x$, these relations exhaust the full set of commutation relations for $\frake_9$.

The affine levels of $\frake_9$ are the levels in the level decomposition with respect to the affine root, corresponding to $e_9$ and $f_9$. It follows from (\ref{derivation}) that the affine level is measured by minus the derivation, 
$-d$, and since $d=-{K^{9}}_{9}$, the affine level is just the number of times 9 appears upstairs minus the number of times it appears downstairs (if we write out all the 9-tuples).
The isomorphism from $\frake_8$ to any subspace in the current algebra is now given by
\begin{align}
Z^i{}_{(p)}&= {E}^{(3p+3)}{}^i{}_9,\nonumber \\
Z^{ijklmn}{}_{(p)}&= {E{}^{(3p+2)}{}^{ijklmn}},\nonumber\\
Z^{ijk}{}_{(p)}&= {E^{(3p+1)}{}^{ijk}}{},\nonumber\\
{G^i}_j{}_{(p)}&= {E^{(3p)}{}^i}_j{}-{\delta^i}_j{E^{(3p)9}}_{9},\nonumber\\
Z_{ijk}{}_{(p)}&= {E^{(3p-1)}{}_{ijk}}{},\nonumber\\
Z_{ijklmn}{}_{(p)}&= {E{}^{(3p-2)}{}_{ijklmn}},\nonumber\\
Z_i{}_{(p)}&= {E^{(3p-3)}{}^9}_i{}, \label{affinisering}
\end{align}
for all integers $p$, together with $c={K^{10}}_{10}$ and $d=-{K^9}_9$. Here $i,\,j,\ldots=1,\,2,\,\ldots,\,8$ and we have dualized the generators so that $9$ appears explicitly only at level $3q+3$ for integers $q$.
The middle equation in (\ref{affinisering}) can equivalently be expressed as
\begin{align}
{E^{(3p)}{}^i}_{j}={G^i}_j{}_{(p)}-\tfrac{1}{9}{\delta^i}_jG^k{}_k{}_{(p)}.
\end{align}
For $p=0$, we set
\begin{align}
{E^{(0)}{}^i}_{j}= {K^i}_j - {\delta^i}_j({K^{a}}_{a}+c),
\end{align}
where $i,\,j,\,\ldots=1,\,2,\,\ldots,\,8$ and ${E^9}_{9 (0)}={K^9}_{9}$.
Thus we have $E^{(3p)}{}^a{}_a =0$ for $p \neq 0$ and $E^{(0)}{}^a{}_a =c$.
\newpage
\noindent
The level decomposition of $\frake_9$ can be illustrated by the following picture:
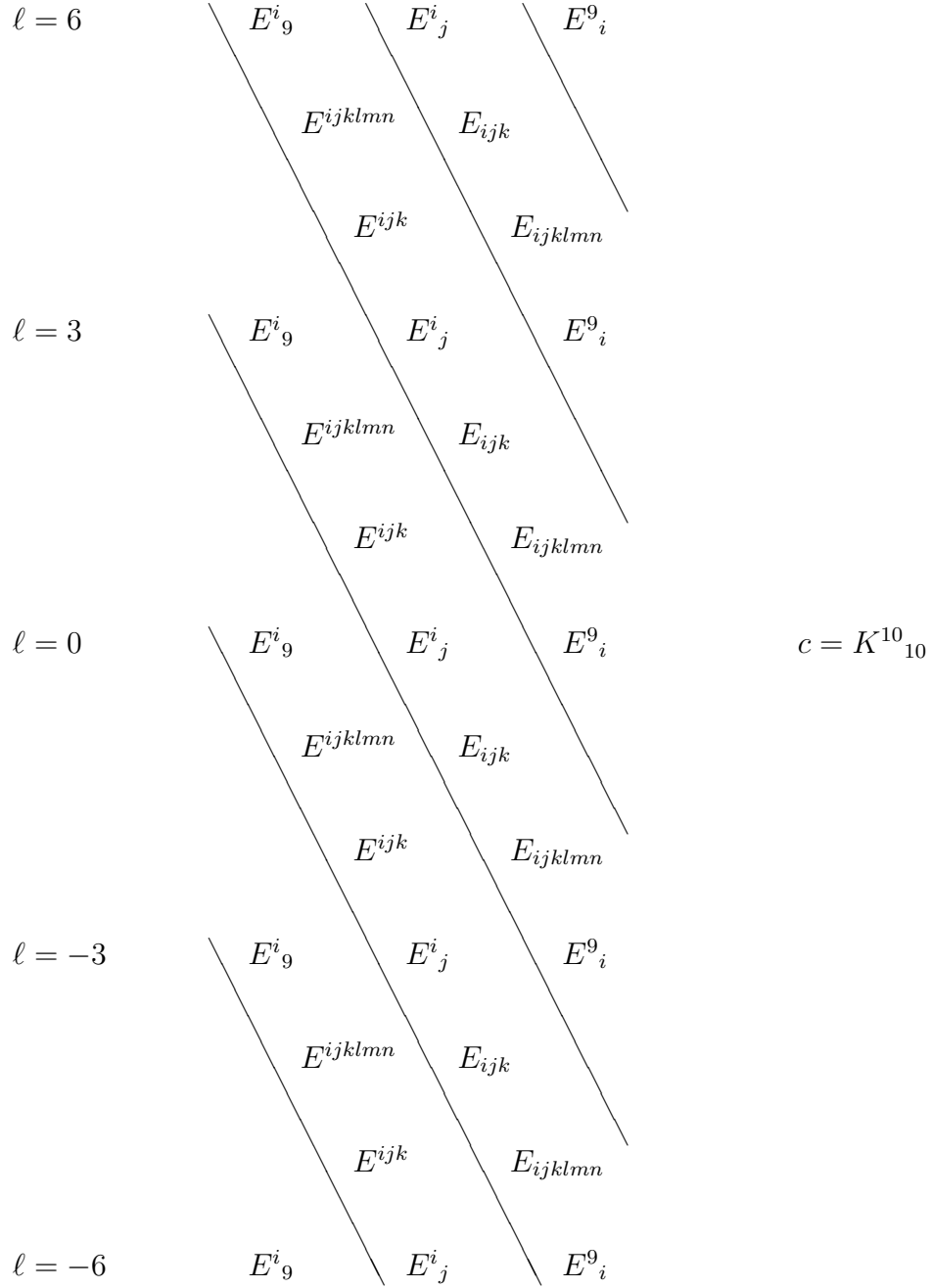
\begin{center}
\scalebox{1}{
\begin{picture}(20,500)
\put(-60,0){${E^i}_9$}
\put(0,0){${E^i}_j$} \put(-150,0){$\ell=-6$}
\put(60,0){${E^9}_i$}
\put(-20,40){$E^{ijk}$}
\put(40,40){$E_{ijklmn}$}
\put(-40,80){$E^{ijklmn}$}
\put(20,80){$E_{ijk}$}
\put(-60,120){${E^i}_9$}
\put(0,120){${E^i}_j$} \put(-150,120){$\ell=-3$}
\put(60,120){${E^9}_i$}
\put(-150,240){$\ell=0$} 
\put(150,240){$c={K^{10}}_{10}$} 
\put(-20,160){$E^{ijk}$}
\put(40,160){$E_{ijklmn}$}
\put(-40,200){$E^{ijklmn}$}
\put(20,200){$E_{ijk}$}
\put(-60,240){${E^i}_9$}
\put(0,240){${E^i}_j$} \put(-150,360){$\ell=3$}
\put(60,240){${E^9}_i$}
\put(-20,280){$E^{ijk}$}
\put(40,280){$E_{ijklmn}$}
\put(-40,320){$E^{ijklmn}$}
\put(20,320){$E_{ijk}$}
\put(-60,360){${E^i}_9$}
\put(0,360){${E^i}_j$} \put(-150,480){$\ell=6$}
\put(60,360){${E^9}_i$}
\put(-20,400){$E^{ijk}$}
\put(40,400){$E_{ijklmn}$}
\put(-40,440){$E^{ijklmn}$}
\put(20,440){$E_{ijk}$}
\put(-60,480){${E^i}_9$}
\put(0,480){${E^i}_j$} 
\put(60,480){${E^9}_i$}
\put(45,490){\line(1,-2){40}}
\put(-15,490){\line(1,-2){100}}
\put(-75,490){\line(1,-2){160}}
\put(-75,370){\line(1,-2){160}}
\put(-75,250){\line(1,-2){127}}
\put(-75,130){\line(1,-2){67}}
\thicklines
\end{picture}}\end{center}
\vspace*{0.4cm}
The horizontal rows correspond to the $\sl(9)$ levels, while the diagonal lines separate the affine levels, with level zero in the middle.
In addition we have the central element $c$ 
at affine level zero.
For better readability, we have suppressed the superscripts $(\ell)$, specifying the $\sl(n)$ levels.

\section{The hyperbolic algebra $\frake_{10}$}
\label{e10sec}

We have finally arrived at $\frake_{10}$, 
the hyperbolic algebra
to which we will attach great importance
in the next chapter when we return to eleven-dimensional supergravity.
This Kac-Moody algebra is also interesting from a mathematical point of view. For example, it has a self-dual root lattice (a feature that it shares with 
$\frake_8$), and all other simply laced hyperbolic Kac-Moody algebras can be
embedded in $\frake_{10}$ \cite{viswanath-2008}.
The Dynkin diagram of $\frake_{10}$ is given below.
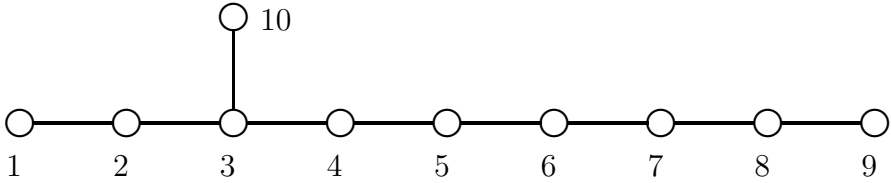
\begin{center}
\scalebox{1}{
\begin{picture}(350,60)
\put(5,-10){$1$}
\put(45,-10){$2$}
\put(85,-10){$3$}
\put(125,-10){$4$}
\put(165,-10){$5$}
\put(205,-10){$6$}
\put(245,-10){$7$}
\put(285,-10){$8$}
\put(325,-10){$9$}
\put(100,45){$10$}
\thicklines
\multiput(10,10)(40,0){9}{\circle{10}}
\multiput(15,10)(40,0){8}{\line(1,0){30}}
\put(90,50){\circle{10}} \put(90,15){\line(0,1){30}}
\end{picture}}\end{center}
\vspace*{0.4cm}
We recall once again that we have the following pattern among the $\sl(n)$ representations that appear at level $\ell>0$ in the grading of $\frake_{10}$ given by the exceptional root (labeled 10 in the Dynkin diagram above):
\begin{alignat}{2}
\ell=3k+3: &\quad E^{\,\cdots\,|a|bcdefghi}&&=E^{\,\cdots\,|a|[bcdefghi]} \nonumber\\
\ell=3k+2: &\quad E^{\,\cdots\,|abcdef}&&=E^{\,\cdots\,|[abcdef]} \nonumber  \\
\ell=3k+1: &\quad E^{\,\cdots\,|abc}&&=E^{\,\cdots\,|[abc]} 
\end{alignat}
where the ellipsis represents $k$ tuples of 9 antisymmetric indices each. Unlike the affine algebra $\frake_9$, for which all representations have this form, $\frake_{10}$ has infinitely many additional basis elements, and the total number grows exponentially with the $\sl(10)$ level. For $\frake_9$ we could replace the 9-tuples with a non-negative integer, specifying the number of tuples. For $\frake_{10}$ we can instead by dualization replace each 9-tuple with an 
index downstairs -- it is enough to specify which of the possible ten values that does not appear among the nine indices.
Thus these representations take the form
\begin{alignat}{2}
\ell=3k+3: &\quad E_{a_1 \cdots a_k}{}^{b|cdefghij}
\nonumber\\
\ell=3k+2: &\quad E_{a_1 \cdots a_k}{}^{bcdefg}
\nonumber  \\
\ell=3k+1: &\quad E_{a_1 \cdots a_k}{}^{bcd}
\end{alignat}
for any $k \geq 0$, symmetric in the $k$ indices downstairs. In the context of the geodesic sigma model based on the coset $E_{10}/K(E_{10})$ that we will review in the next chapter, these representations have been conjectured to
correspond to spatial derivatives of the fields in eleven-dimensional supergravity at a fixed spatial point. This is a natural guess, since the dynamical equivalence between the supergravity theory and the $E_{10}$ model,
which we will consider in the next chapter, 
only holds at a fixed spatial point.
The supergravity theory is then truncated to the lowest order spatial derivatives, and the $E_{10}$ model is truncated to the lowest levels in the level decomposition.
Thus adding symmetric indices would somehow correspond to acting with (commuting) partial derivatives.

\subsection{Decomposition under $\mathfrak{e}{}_8 \oplus \mathfrak{a}{}_1$}
\label{e10frome8}
Before studying the $E_{10}$ coset model in the $\sl(10)$ decomposition, we 
consider the decomposition under $\mathfrak{e}{}_8 \oplus \mathfrak{a}{}_1$, where the appearing representations can be interpreted differently.

In section \ref{affinechapter} we saw that the Lie algebra $\frake_8$ is also a generalized Jordan triple system with the triple product
\begin{align}
(xyz)=[[x,\,y],\,z].
\end{align}
Furthermore, we said that this triple system is isomorphic to the generalized Jordan triple system
$(\frake_9){}_{-1}$. 
In section \ref{extensions} we reviewed Theorem 2.1
in {\gcrkma}, which easily can be generalized to affine Kac-Moody algebras 
$\mathfrak{h}$. Applied to the case $\mathfrak{h}=\frake_9$
it says 
that the triple system
$(\frake_9){}_{-1}{}^{(2)}$ is isomorphic to $(\frake_{10}){}_{-1}$, where the grading of $\g=\frake_{10}$ is given by the node labeled 8 in the Dynkin diagram above. Thus we can conclude that the triple system
$\frake_8{}^{(2)}$
is isomorphic to the generalized triple system $(\frake_{10}){}_{-1}$, derived from $\frake_{10}$ with this grading.
It follows that the structure constants of the triple system 
$(\frake_{10})_{-1}$ can be written 
\begin{align}
f^{\vA}{}_a{}^\vB{}_b{}^\vC{}_c{}^{\vD}{}_{d}{}=
f^{\vA\vB\vC\vD}\de_{ab}\de_{cd}
-\eta^{\vA\vB} \eta^{\vC\vD} \de_{ab}\de_{cd}
+\eta^{\vA\vB}\eta^{\vC\vD}
\de_{bc}\de_{ad},
\end{align}
where $f^{\vA\vB\vC\vD}=f^{\vA\vB}{}_\vE f^{\vE\vC\vD}$. Furthermore, 
$f^{\vA\vB}{}_\vC$ and $\eta^{\vA\vB}$ are the structure constants and the components of the Killing form of $\frake_8$, as in section \ref{e8section}.
It is clear that the representation at the first level in the $\frake_8 \oplus \mathfrak{a}{}_1$ decomposition is $({\bf248},\,{\bf2})$. Continuing to level two, we have
\begin{align}
g^{\vA}{}_a{}^\vB{}_b{}^\vC{}_c{}^{\vD}{}_{d}{}=
f^{\vA}{}_a{}^\vC{}_c{}^\vB{}_b{}^{\vD}{}_{d}{}-
f^{\vB}{}_b{}^\vC{}_c{}^\vA{}_a{}^{\vD}{}_{d}{}
\end{align}
and comparing with the expressions (\ref{e8projektorer}) for the $\frake_8$ projectors for the irreducible parts of ${\bf 248} \times {\bf 248}$, we get
\begin{align}
-\tfrac14g^{\vA}{}_a{}^\vB{}_b{}^\vC{}_c{}^{\vD}{}_{d}{}&=
7 (\mathbb{P}{}_{\bf3875})^{\vA\vB\vC\vD}(\mathbb{P}{}_{\bf1}){}_{abcd}\nn\\&\quad\,
+15 (\mathbb{P}{}_{\bf248})^{\vA\vB\vC\vD}(\mathbb{P}{}_{\bf3}){}_{abcd}\nn\\&\quad\,
+31 (\mathbb{P}{}_{\bf1})^{\vA\vB\vC\vD}(\mathbb{P}{}_{\bf1}){}_{abcd},
\end{align}
where $(\mathbb{P}{}_{\bf1})^{ab}{}_{cd}=\de^{a}{}_{[c}\de^b{}_{d]}$ and
$(\mathbb{P}{}_{\bf3})^{ab}{}_{cd}=\de^{a}{}_{(c}\de^b{}_{d)}$.
Thus the representations at level two are
$({\bf3875},\,{\bf1})$, $({\bf248},\,{\bf3})$ and $({\bf1},\,{\bf1})$.
In {\gauging} we associate the representations $({\bf3875},\,{\bf1})$ and $({\bf1},\,{\bf1})$
to the irreducible parts of the embedding tensor appearing in gauged supergravity.

\chapter{The $E_{10}$ coset model and maximal supergravity} \label{e10modelchapter}

In chapter \ref{hiddenchapter} we saw that the scalar part of eleven-dimensional supergravity reduced to $d$ dimensions on an $n$-torus 
can be described by a sigma model based on the coset $G/K(G)$. For $4 \leq n \leq 8$, the Lie algebra of the global symmetry group $G$ is $\mathfrak{e}{}_{n(n)}$, the split real form of the complex Lie algebra $\frake_n$.
In the preceding chapter we studied these algebras, not only for $4 \leq n \leq 8$, but also for $n=9$ and $n=10$. 
In this chapter we will, following \cite{Damour:2002cu,Damour:2004zy}, construct a one-dimensional geodesic sigma model based on the infinite-dimensional coset $E_{10}/K(E_{10})$. Although $E_{10}$ is not well understood as a Lie group we can describe the sigma model by the properties of the Lie algebra $\frake_{10}$ that we studied in the preceding chapter. 
In the end of this chapter we will see that there is a correspondence, up to truncations on both sides, between the equations of motion of eleven-dimensional supergravity on the one side, and the dynamics of the $E_{10}$ coset model on the other
\cite{Damour:2002cu,Damour:2004zy}.

\section{Lagrangian}

We start with an $E_{10}$ group element $\vV$,
depending on a parameter $t$,
from which we derive the Maurer-Cartan form
$v=\mathcal{V}^{-1}\partial\mathcal{V}$
where $\pa$ denotes derivative with respect to $t$.
Let $\vQ$ and $\vP$ be the projections of the $\frake_{10}$ algebra element $v$ on the 
maximal compact
subalgebra $\mathfrak{k}(\frake_{10})$ and 
the coset $\mathfrak{p}$,
respectively:
\begin{align}
v&=\vP+\vQ,&
\vP &=\tfrac12(v - \omega(v)),&
\vQ &= \tfrac12(v + \omega(v)),
\end{align}
These projections correspond to the symmetric and antisymmetric parts of the 
$\sl(n)$ element that we considered in (\ref{uppdelning}). The expression 
\begin{align}
{P}{}_\al{}_{ab} {P}{}^\al{}_{ab}=\tr(P_\al P^\al)
\end{align}
that appeared in the Lagrangian (\ref{enkellagrangian}), can be written
$\kappa(P_\al,\,P^\al)$ if we consider $P_\al$ as an element in $\sl(n)$.
Thus for the $E_{10}$ model we consider analogously the Lagrangian
\begin{align} \label{e10lagrange}
\vL&= 
\tfrac{1}{4}n^{-1} \kappa( \vP ,\, \vP ),
\end{align}
where $n$ is a Lagrange multiplier needed for reparametrization invariance along the geodesic (and $1/4$ is a convention).
We expand $\vP$ and $\vQ$ in the $\sl(10)$ bases of $\frakk(\frake_{10})$ that we described in the preceding chapter,
\begin{align}
\vP&=P_{ab}S^{ab}
+P_{abc}S^{abc}+P_{abcdef}S^{abcdef}+\cdots
,\nn\\
\vQ&=Q_{ab}J^{ab}
+Q_{abc}J^{abc}+Q_{abcdef}J^{abcdef}+\cdots.
\end{align} 
In the same way as we explained for $SL(n)/SO(n)$ in chapter 
\ref{hiddenchapter}, the Lagrangian (\ref{e10lagrange}) has a global $E_{10}$ symmetry and a 
local $K(E_{10})$ symmetry. 
We can use the local $K(E_{10})$ symmetry to choose 
$v$ to always be an element in the subalgebra
\begin{align}
(\frake_{10})_0 + (\frake_{10})_+ = (\frake_{10})_0 + (\frake_{10})_1
+ (\frake_{10})_2 + \cdots
\end{align}
of $\frake_{10}$, where the grading is given by the exceptional root. 
Thus $\vP$ and $\vQ$ have the same components in this gauge, except at level zero,
\begin{align} \label{borelgaugen}
\vP&=P_{ab}S^{ab}
+P_{abc}S^{abc}+P_{abcdef}S^{abcdef}+\cdots
,\nn\\
\vQ&=Q_{ab}J^{ab}
+P_{abc}J^{abc}+P_{abcdef}J^{abcdef}+\cdots.
\end{align} 
The group element $\vV$ can furthermore be written as
$\mathcal{V}=e^X 
e^{h}$
where $h$ and $X$ are algebra elements, expanded in the basis as
$h=h_a{}^bK^a{}_b$
at level zero, and
\begin{align}
X&=
X_{abc}E^{abc}+X_{abcdef}E^{abcdef}+\cdots
\end{align}
at higher levels.
We then have
\begin{align}
e^{-h} e^{-X} \partial (e^X e^{h})
= e^{-h} e^{-X} \partial e^X e^{h}
+ e^{-h} \partial e^{h}.
\end{align}
By the Baker-Campbell-Hausdorff formula we get
\begin{align}
e^{-X}\pa e^X &= \pa X +\tfrac12 [\pa X,\,X]+\cdots\nn\\&
=\pa X_{abc}E^{abc}+
(\pa X_{abcdef}-\tfrac12 X_{abc}\pa X_{def}) E^{abcdef}+\cdots\nn\\&
=D X_{abc}E^{abc}+
D X_{abcdef}E^{abcdef}+\cdots,
\end{align}
where we have defined the `covariant derivatives' \cite{Damour:2002cu}
\begin{align}
D X_{abc}&=\pa X_{abc},\nn\\
D X_{abcdef}&=\pa X_{abcdef}-\tfrac12 X_{abc}\pa X_{def}.
\end{align}
It follows from (\ref{borelgaugen}) that
the sigma model Lagrangian truncated to the first two positive levels is given by 
\begin{align}
\mathcal{L} 
\label{lagrange1}
&=
n^{-1}(P_{ab}P_{ab}-P_{aa}P_{bb})
\nn\\&\quad\,
+ \tfrac{1}{2}n^{-1}
P_{abc}P_{def}\kappa( E^{abc} ,\,  F_{def} )\nn\\&\quad\,
+ \tfrac{1}{2}n^{-1} P_{abcdef}P_{ghijkl}\kappa( E^{abcdef} ,\,  F_{ghijkl} ) +\cdots.
\end{align}
We can write the components of $\vP$ and $\vQ$ at level zero as
\begin{align}
P_{ab}&=\tfrac{1}{4}(e_a{}^m\pa{e}{}_m{}^b+e_b{}^m\pa{e}{}_m{}^a),
&
Q_{ab}&=\tfrac{1}{4}(e_a{}^m\pa{e}{}_m{}^b-e_b{}^m\pa{e}{}_m{}^a),
\end{align}
where $e_m{}^a$ are the components of the $GL(10)$ matrix $e^{h}$ and 
$e_a{}^m$ those of the inverse $e^{-h}$. We can consider $e_a{}^m$ as a `vielbein' and 
accordingly 
introduce the `metric'
\begin{align}
g_{mn}&=e_m{}^a e_n{}^a, & g^{mn}&=e_a{}^m e_a{}^n.
\end{align}
We then get
\begin{align}
\kappa( e^{-h} \pa e^{h} + (e^{-h} \pa e^{h})^t ,\,
e^{-h} \pa e^{h} + (e^{-h} \pa e^{h})^t )
=\pa{g}_{mn}\pa{g}_{pq}({g}^{mp}{g}^{nq}-{g}^{mn}{g}^{pq})
\end{align}
and the truncated Lagrangian takes the form
\begin{align}
\label{lagrange4}
\mathcal{L}&=\mathcal{L}_0 + \tfrac{1}{2}n^{-1}
(3! DX_{mnp}DX^{mnp}
+6!DX_{mnpqrs}DX^{mnpqrs}
+\cdots),
\end{align}
where the indices are raised with 
$g^{mn}$, and
\begin{align}
\mathcal{L}_0
&=
n^{-1}(P_{ab}P_{ab}-P_{aa}P_{bb})\nn\\
&=n^{-1} 
(\tfrac18 v_a{}^b v_a{}^b + \tfrac18 v_b{}^a v_a{}^b 
-\tfrac14 v_a{}^a v_b{}^b\nn\\
&=
\tfrac{1}{16}n^{-1}\pa{g}_{mn}\pa{g}_{pq}({g}^{mp}{g}^{nq}-{g}^{mn}{g}^{pq}),
\end{align}
where $v_a{}^b$ are the components of $v$ at level zero, in the $K^a{}_b$ basis
of $\gl(n)$.

\section{Equations of motion}

The variation of the Maurer-Cartan form is
\begin{align}
\de(\vV^{-1} \pa \vV ) &= (\de \vV^{-1})\pa\vV+\vV^{-1}\pa(\de\vV)\nn\\
&= -(\vV^{-1}\de\vV)(\vV^{-1}\pa\vV)-\vV^{-1}\pa(\vV\de\vV^{-1}\vV)\nn\\
&= [\vV^{-1}\pa\vV,\,\vV^{-1}\de\vV]+\pa(\vV^{-1}\de\vV).
\end{align}
Writing
$\vV^{-1}\de \vV =p+q$,
where $q$ and $p$ belong to $\frakk(\frake_{10})$ and the coset, respectively,
and using the invariance of the Killing form, we get
\begin{align}
n^{-1} \kappa( \de \vP,\,\vP)
&= n^{-1} \kappa(\pa p+[\vQ,\,p]-[q,\,\vP],\,\vP)\nn\\
&= n^{-1} \kappa( \pa p,\,\vP)-\kappa( [\vQ,\,\vP],\,p)-\kappa([\vP,\,\vP],\,q)\nn\\
&= - \kappa( p,\,\pa (n^{-1}\vP))-n^{-1}\kappa( p,\,[\vQ,\,\vP])\nn\\
&= - \kappa( p,\,\pa (n^{-1}\vP)+n^{-1}[\vQ,\,\vP])
\end{align}
up to a total derivative. 
This gives
\begin{align}
\de (n^{-1}\kappa( \vP ,\,\vP )) 
&= -n^{-2} \de n \kappa (\vP,\,\vP)-2 \kappa( p,\,\pa (n^{-1}\vP)+n^{-1}[\vQ,\,\vP])
\end{align}
(again up to a total derivative) and the equations of motion are
\begin{align}
\kappa ( \vP ,\, \vP )&=0, & n \pa (n^{-1} \vP) + [\vQ,\,\vP]&=0,
\end{align}
where the first equation is the Hamilton constraint
which ensures that the motion is lightlike.
Using the commutation relations for $\frake_{10}$ the other equations can be written
\begin{align} \label{e10eom}
n\pa (n^{-1} P_{ab})&=4Q_{c(a}P_{b)c}-\tfrac14 (P_{acd}P_{bcd}
-\tfrac1{9}\de_{ab}P_{cde}P_{cde})\nn\\&\quad\,
-\tfrac1{240}(P_{acdefg}P_{bcdefg}-\tfrac19\de_{ab}P{}_{cdefgh}P{}_{cdefgh}),\nn\\
n\pa (n^{-1} P_{abc})&=-\tfrac16 P_{abcdef}P_{def}
+6(P_{ad}-Q_{ad})P_{dbc},\nn\\
n\pa (n^{-1} P_{abcdef})&=12(P_{ag}-Q_{ag})P{}_{gbcdef}.
\end{align}

\section{Comparison to eleven-dimensional supergravity}
We will now go back to the equations of motion for eleven-dimensional supergravity that we gave in the first chapter, and study them under assumptions that make them comparable with the equations of motion from the $E_{10}$ model. For this we have to break spacetime covariance and treat space and time separately. We split the curved indices as $M=(t,\,m)$ and the flat indices as $A=(0,\,a)$, where $m,\,n\ldots$ and $a,\,b,\ldots$ denote the ten spatial directions. The signature is mostly plus, $(-+\cdots+)$. We split the elfbein $E_M{}^A$ into a spatial zehnbein $e_m{}^a$ and a lapse function $N$ as
\begin{align}
E^{M}{}_A = 
\begin{pmatrix}
N & 0 \\
0 & e^{m}{}_a
\end{pmatrix}
\end{align}
so that we get
\begin{align}
\Omega_{0ab}&=N^{-1}e_a{}^m\pa_t e_m{}^b, & 
\Omega_{a00}&=e_a{}^mN^{-1}\pa_m N, & 
\Omega_{ab0}&=0.
\end{align}
Following \cite{Damour:2002cu}, we neglect spatial derivatives of second order and higher, and set 
\begin{align}
\omega_{abc}=\omega_{00c}=\omega_{aa0}=0.
\end{align}
One can show that this indeed is a consistent truncation \cite{Damour:2002cu,Damour:2004zy}.
For the remaining components of the spin connection we get
\begin{align}
\omega_{0ab}&=\Omega_{0[ab]},&
\omega_{00a}&=\Omega_{a00},&
\omega_{ab0}&=\Omega_{0(ab)}.
\end{align}
Under these assumptions, the equations of motion become
\begin{align} \label{sugraeom}
n\pa_t (n^{-1} \omega_{abt})&=2\omega_{tc(a}\omega_{b)ct}-\tfrac12 (F_{tacd}F_{tbcd}
-\tfrac1{9}\de_{ab}F_{tcde}F_{tcde})\nn\\&\quad\,
-\tfrac1{240}(\hat{F}{}_{tacdefg}\hat{F}{}_{tbcdefg}-\tfrac19\de_{ab}\hat{F}{}_{tcdefgh}\hat{F}{}_{tcdefgh}),\nn\\
n\pa_t (n^{-1} F_{tabc})&=-\tfrac16 \hat{F}{}_{tabcdef}F_{tdef}
+3(\omega_{adt}-\omega_{tad})F_{tdbc},\nn\\
n\pa_t (n^{-1} \hat{F}{}_{tabcdef})&=6(\omega_{agt}-\omega_{tag})\hat{F}{}_{tgbcdef}.
\end{align}
where we have dualized the field strength $F$ to
\begin{align}
\hat{F}{}^{ABCDEFG}=\tfrac1{4!}\ep^{ABCDEFGHIJK}F_{HIJK}.
\end{align}
The equation of motion for the dual field strength $\hat{F}$ has been computed by dualizing the Bianchi identity for $F$.
We see now that the equations (\ref{e10eom}) coincide with (\ref{sugraeom}) if we set
\begin{align}
2P_{ab}(t)&=\omega_{abt}(t,{\bf x}_0), & P_{abc}(t)&=F_{tabc}(t,{\bf x}_0), \nn\\
2Q_{ab}(t)&=\omega_{tab}(t,{\bf x}_0),&
P_{abcdef}(t)&=\hat{F}{}_{tabcdef}(t,{\bf x}_0),
\end{align}
where ${\bf x}_0$ is a fixed, but arbitrarily chosen, spatial point.

The analysis has been carried out in detail in \cite{Damour:2002cu,Damour:2004zy}
for the first three positive levels, including also the spatial components of the spin connection on the supergravity side. However, at 
level three the first mismatches appear, and at higher levels it is not clear how to interpret the representations.
As we mentioned in section \ref{e10sec}, some of them can be interpreted as spatial derivatives, but there is also a number of additional representations,
which grows exponentially for each level. In conclusion, the $E_{10}$ model has certainly proved successful to some extent, but much more work remains to make the picture clear.
We recommend the reviews \cite{Damour:2002et,Kleinschmidt-nicolai-revy,Henneaux:2007ej} for further reading.

\clearpage
\pagestyle{plain}
\def\href#1#2{#2}
\addcontentsline{toc}{chapter}{\sffamily\bfseries Bibliography}


%
%





\providecommand{\href}[2]{#2}\begingroup\raggedright\endgroup

\end{document}